\documentclass[twocolumn,showpacs,floatfix,superscriptaddress]{revtex4}
\usepackage{latexsym}
\usepackage{graphics}
\usepackage{epsf}
\usepackage{epsfig}
\usepackage{graphicx}
\usepackage{dcolumn}
\usepackage{bm}
\usepackage{latexsym}
\usepackage{booktabs}
\usepackage{amssymb}
\usepackage[english]{babel}
\usepackage{hyperref}
\usepackage{amsfonts}

\begin{document}

\noindent 

\title{Reducing Frustration in Spin Systems: Social Balance as an XOR-SAT problem}

\author{Filippo Radicchi}\email[]{f.radicchi@iu-bremen.de}
\affiliation{School of Engineering and Science , International University Bremen , P.O.Box 750561 , D-28725 Bremen , Germany.}
\author{Daniele Vilone}\email[]{d.vilone@iu-bremen.de}
\affiliation{School of Engineering and Science , International University Bremen , P.O.Box 750561 , D-28725 Bremen , Germany.}
\author{Sooeyon Yoon}\email[]{syyun95@gmail.com}
\affiliation{Department of Physics and Research Institute of Basic Sciences , Kyung Hee University , Seoul 130-701 , Korea.}
\author{Hildegard Meyer-Ortmanns}\email[]{h.ortmanns@iu-bremen.de}
\affiliation{School of Engineering and Science , International University Bremen , P.O.Box 750561 , D-28725 Bremen , Germany.}



\begin{abstract}
Reduction of frustration was the driving force in an approach to
social balance as it was recently considered by Antal \emph{et al.} [ T. Antal, P. L. Krapivsky, and S. Redner , Phys. Rev. E {\bf 72} , 036121 (2005). ]. We generalize their triad
dynamics to $k$-cycle dynamics for arbitrary integer $k$. We
derive the phase structure, determine the stationary solutions and
calculate the time it takes to reach a frozen state. The main
difference in the phase structure as a function of $k$ is related
to $k$ being even or odd. As a second generalization we dilute the
all-to-all coupling as considered by Antal \emph{et al.} to a random
network with connection probability $w<1$. Interestingly, this
model can be mapped onto a $k$-XOR-SAT problem that is studied in
connection with optimization problems in computer science. What is
the phase of social balance in our original interpretation is the
phase of satisfaction of all clauses without frustration in the
satisfiability problem of computer science. Nevertheless, although
the ideal solution without frustration always exists in the cases
we study, it does not mean that it is ever reached, neither  in
the society nor in the optimization problem, because the local
dynamical updating rules may be such that the ideal state is
reached in a time that grows exponentially with the system size.
We generalize the random local algorithm usually applied for
solving the $k$-XOR-SAT problem to a $p$-random local  algorithm,
including a parameter $p$, that corresponds to the propensity
parameter in the social balance problem. The qualitative effect is
a bias towards the optimal solution and a reduction of the needed
simulation time. We establish the mapping between the $k$-cycle
dynamics for social balance on diluted networks and the
$k$-XOR-SAT problem solved by a $p$-random local algorithm.
\end{abstract}


\maketitle

\section{Introduction}
\label{intro_sec}
Recently Antal \emph{et al.} \cite{antal} proposed a triad dynamics to
model the approach of social balance. An essential ingredient in
the algorithm is the reduction of frustration in the following
sense. To an edge (or link) in the all-to-all topology is assigned a
value of $+1$ or $-1$ if it connects two individuals who are friends
or enemies respectively. The sign $\pm 1$ of a link we call also
its spin. If the product of links along the boundary of a triad is
negative, the triad is called frustrated (or imbalanced), otherwise it is called
balanced (or unfrustrated). The state of the network is called balanced if all
triads are balanced. If the balanced state is achieved by all
links being positive the state is called ``paradise''. The algorithm depends on a parameter $p \in [0,1]$ 
called propensity which determines the tendency of the system to
reduce frustration via flipping a negative link to a positive one
with probability $p$ or via flipping a positive link to a negative with
probability $1-p$. For an all-to-all topology Antal \emph{et al.} predict a transition from imbalanced stationary
states for $p<1/2$ to balanced stationary
states for $p\geq 1/2$. Here the dynamics is motivated by social
applications so that the notion of frustration from physics goes
along with frustration in the psychological sense.
\\
Beyond frustration in social systems, within physics, the notion
is familiar from spin glasses. It is the degree of frustration in
spin glasses which determines the qualitative features of the
energy landscape. A high [low] degree of frustration corresponds
to many [few] local minima in the energy landscape.
In terms of energy landscape it was speculated by Sasai and Wolynes \cite{wolynes} that is the low degree of frustration in a
genetic network which is responsible for the few stable cell
states in the high-dimensional space of states.
\\
Calculational tools from spin-glass theory like the replica-method \cite{parisi}
turned out to be quite useful in connection with generic
optimization problems (as they occur, for example, in computer
science) whenever there is a map between the spin-glass
Hamiltonian and a cost function. The goal in finding the ground
state-energy of the Hamiltonian translates to the minimization of
the costs. A particular class of the optimization problems refers
to the satisfiability problems. More specifically one has a system
of $B$ Boolean variables and $Q$ logical constraints (clauses)
between them. In this case, minimizing the costs means minimizing
the number of violated constraints. In case of the existence of a non-violating configuration the
problem is said to be satisfiable, it has a zero-ground state
energy in the Hamiltonian language. Here it is obvious that
computer algorithms designed to find the optimal solution have to
reduce the frustration down to a minimal value. So the reduction
of frustration is in common to very different dynamical processes.
\\
The algorithms we have to deal with belong to the so-called
incomplete algorithms \cite{garey,weigt,semerjian} characterized by some kind
of Monte-Carlo dynamics that tries to find the solution via
stochastic local moves in configuration space, starting from a
random initial configuration. It either finds the solution "fast"
or never (this will be made more precise below). Among the
satisfiability problems there are the $k$-SAT ($k$S) problems \cite{cook,mezard,mezard2}, for
which actually no frustration-free solution exists above a certain
threshold in the density of clauses imposed on the system. In this
case the unsatisfiability is not a feature of the algorithm but
intrinsic to the problem. However, there is a special case of
$k$S problems, so-called $k$-XOR-SAT ($k$XS) problems \cite{weigt,semerjian,mezard2,cocco} which are
always solvable by some global algorithm, but poses a challenge for
finding the solution by some kind of Monte-Carlo dynamics, very
similar to the one used for solving the $k$S problem, where actually no solution may exist. Now it is these $k$XS problems and their solutions that
are related to the social balance dynamics.
\\
In particular it can be easily shown \cite{mezard,mezard2,cocco} that the
satisfiability problem $3$S (and also the subclass $3$XS) can first be mapped onto a
$3$-spin model that is a spin-glass, and as we shall show below,
the $3$-spin glass model can next be mapped onto the triad
dynamics of Antal \emph{et al.} \cite{antal}. The $k$XS problem is usually
studied for diluted connections because the interesting changes in
the phase structure of the $k$XS problem appear at certain
threshold parameters in the dilution, while the all-to-all case is
not of particular interest there.
\\
Dilution of the all-to-all topology is not only needed for the
mapping to the $3$XS problem in its usual form. It is also a
natural generalization of the triad dynamics considered in
\cite{antal} for social balance. A diluted network is more
realistic than an all-to-all topology by two reasons: either two
individuals may not know each other at all (this is very likely
in case of a large population size) or they neither like or
dislike each other, but are indifferent as emphasized in
\cite{cartwright} as an argument for the postulated absence of
links. For introducing dilution into the all-to-all network
considered by Antal \emph{et al.} it is quite natural to study random
Erd\"os-R\'enyi networks \cite{erdos} for which two nodes are connected by a link with probability $w$. On the other hand, dilution in the
$k$XS problem is parameterized by the ratio $\alpha$ of
number of clauses over number of variables (variables in the corresponding
spin model or number of links in the triad dynamics). We will
determine the map between both parameterizations.
\\
\vspace{5pt}
\\
In the first part of this paper (section \ref{model}) we generalize the triad dynamics
to $k$-cycle dynamics, driven by the reduction of frustration,
with arbitrary integer $k$. In the context
of \emph{social balance} theory, Cartwright and Harary \cite{cartwright}
introduced the notion of balance describing a balanced state
with all $k$-cycles  being balanced and $k$ not
restricted to three. We first study such model on fully connected networks (section \ref{complete}). For given fixed and integer $k\geq 3$ in
the updating rules, we draw the differential equations of the time evolution due to the local dynamics (section \ref{evolution}) and we predict the stationary densities of
$k$-cycles, $k$ arbitrary integer, containing $j \leq k$ negative
links (section \ref{stationary}). As long as $k$ is odd (section \ref{odd}) in the updating dynamics, the results
are only quantitatively different from the case of $k=3$
considered in \cite{antal}. An odd cycle of length three is,
however, not an allowed loop in a bipartite graph, for which links
may only exist between different type of vertices so that the
length of a loop of minimal size in a bipartite graph is four. In
addition, a $4$-cycle with four negative links (that is four
individuals each of which dislikes two others) is balanced and not
frustrated, although it may be called the ``hell'', so it does not
need to be updated in order to reduce its frustration. (To call the hell
with four negative links balanced is not specific for the notion
of frustration in physics; also in social balance theory it is the
product over links in the loop which counts and decides about
balance or frustration \cite{roberts}.) This difference is
essential as compared to the triad dynamics, in which a triad of
three unfriendly links is always updated. It has important
implications on the phase structure as we will show. For even values of $k$
and larger than four, again there are only quantitative
differences in the phase structure as compared to $k=4$ (section \ref{even}).
\\
As in \cite{antal} , for odd values of $k$, we shall distinguish between stationary states
in the infinite volume limit that can be either balanced (for
$p \geq 1/2$) or frustrated (for $p< 1/2$) since it is not
possible to reach the paradise in a finite time. They are
predicted as solutions of mean field equations. In numerical
simulations, fluctuations about their stationary values do not die out
in the phase for $p< 1/2$ so that some frustration remains,
while for $p \geq 1/2$ frozen states are always reached in the
form of the paradise although other balanced states with a finite
amount of negative links are in principle available, but are quite
unlikely to be realized during the finite simulation time. They
are exponentially suppressed due to their small weight in
configuration space. We calculate the time it takes to reach a
frozen state at and above the phase transition (section \ref{time_odd}). For even values of $k$ we
have only two types of stationary frozen states, ``paradise'' and
``hell'' with all links being positive and negative, respectively.
In this case the time to reach the frozen states at the transition
can be calculated in two ways. The first possibility applies for
both even and odd values of $k$ and is based on calculating the time
it takes until a fluctuation is of the same order in size as the
average density of unfriendly links. The second one, applicable to the case of even values of $k$, can be
obtained by mapping the social system to a Markov process known as
the Wright-Fisher model for diploid organisms \cite{wright}, for which the decay time to one of the final configurations (all ``positive'' or all
``negative'' genes) increases quadratically in the size $N$ of the
system (section \ref{sec:time_even}).
\\
In the second part we generalize the $k$-cycle dynamics to diluted
systems (section \ref{sec:diluted}). The dilution, originally given in terms of the
probability for connecting two links in a random Erd\"os-R\'enyi
network \cite{erdos}, is then parameterized in terms of the dilution parameter $\alpha$, and the
results for stationary and frozen states and the time needed to
reach them will be given as a function of $\alpha$ (section \ref{sec:ratio}). The original
triad dynamics of Antal \emph{et al.} with propensity parameter $p$ on a
diluted network contains, as special case,  the usual
Random-Walk SAT (RWS) algorithm for finding the solution of the
$3$XS problem corresponding to the choice of $p=1/3$ in
the triad dynamics. Therefore it is natural to generalize the RWS algorithm  for generic $p\in [0,1]$ and to study the
modifications in the performance of the algorithm as a function of
$p$ (section \ref{sec:RWS}). For the $k$S problem, and similarly for the
$k$XS problem, there are three thresholds in $\alpha$,
$\alpha_d$, $\alpha_s$, and  $\alpha_c$ with
$\alpha_d<\alpha_s<\alpha_c$. Roughly speaking, the threshold
$\alpha_d$ corresponds to a dynamical transition between a phase
in which the RWS algorithm finds a solution in a time
linearly increasing with the size of the system for
$\alpha<\alpha_d$, and exponentially increasing with the system
size for $\alpha>\alpha_d$. The value $\alpha_s$ characterizes a
transition in the structure of the solution space, from one
cluster of exponentially many solutions ($\alpha<\alpha_s$) to
exponentially many clusters of solutions ($\alpha>\alpha_s$).
Finally, $\alpha_c$ refers to the transition between satisfiable
and unsatisfiable $k$S problems, this means that for these
models not all constraints can be satisfied simultaneously in the
UNSAT-phase for $\alpha>\alpha_c$ so that a finite amount of
frustration remains. Above this last threshold lies a value of
$\alpha,\; \alpha=\alpha_m$, such that for $\alpha>\alpha_m$ the
mean field approximation is justified that was used for the maximum value of $\alpha$ in the all-to-all topology of the triad dynamics of
\cite{antal}. We shall study the influence of the parameter $p$ on
the value of $\alpha_d$ (section \ref{sec:alphad}) and on the Hamming distance for $\alpha$
smaller $\alpha_s$ or larger $\alpha_s$ (section \ref{sec:alphas}). Moreover we will show how the choice of $p$ changes the possibility to find a solution for the $k$XS problem (section \ref{sec:pc}) and we will  determine the validity
range of the mean-field approximation  (section \ref{sec:alpham}). As it turns out, the
parameter $p$ introduces some bias in the RWS,
accelerating the convergence to ``paradise'' and reducing the
explored part of configuration space. On the other hand, an
inappropriate choice of $p$ or too much dilution may prevent an
approach to paradise.
Fluctuations in the wrong direction, increasing the amount of
frustration, go along with improved convergence to the balanced
state.


\section{The Model for Social Balance}
\label{model} 
\begin{figure}[ht]
\includegraphics*[width=0.47\textwidth]{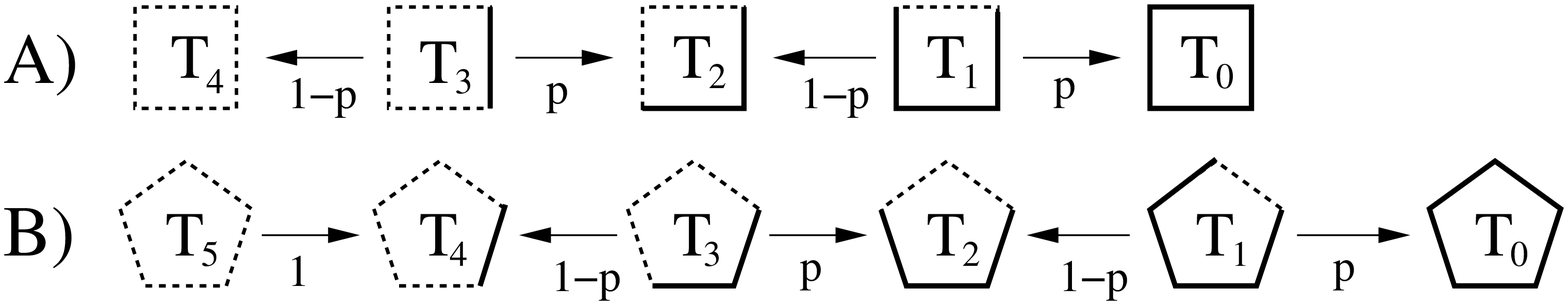}
\caption{Dynamical rules in case of $k=4$ (A) and $k=5$ (B). The
cycles containing an odd number of ``unfriendly'' links are
considered as imbalanced and evolve into balanced ones. Full and
dashed lines represent ``friendly'' and ``unfriendly'' links
respectively.} \label{fig:example}
\end{figure}

We represent individuals as vertices (or nodes) of a
graph and a relationship between two individuals as a link (or
edge) that connects the corresponding vertices. Moreover we assign
to a link $(i,j)$ between two nodes $i$ and $j$ a binary spin
variable $s_{i,j}=\pm 1$, with $s_{i,j}=1$ if the individuals $i$
and $j$ are friends , and $s_{i,j}=-1$ if  $i$ and $j$ are
enemies. We consider the standard notion of \emph{social balance}
extended to cycles of order $k$ \cite{cartwright,heider}. In
particular a cycle of order $k$ (or a $k$-cycle) is defined as a
closed path between  $k$ distinct nodes $i_1$, $i_2$, \ldots ,
$i_k$ of the network, where the path is performed along the links
of the network $(i_1,i_2)$ , $(i_2,i_3)$ , \ldots ,
$(i_{k-1},i_k)$, $(i_k,i_1)$. Given a value of $k$ we have $k+1$
different types $T_0$, $T_1$, \ldots, $T_j$, \ldots, $T_k$  of
cycles of order $k$ containing $0$, $1$, \ldots, $j$, \ldots, $k$
negative links, respectively. A cycle of order $k$ in the network
is considered as balanced if the product of the signs of links
along the cycle equals $1$, otherwise the cycle is considered as
imbalanced or frustrated. Accordingly, the network is considered
as balanced if each $k$-cycle of the network is balanced.
\\
We consider our social network as a dynamical system. We perform a
local unconstrained dynamics obtained by a natural generalization
of  the local triads dynamics, recently proposed by Antal \emph{et
al.} \cite{antal}. We first fix a value of $k$. Next, at each
update we choose at random  a $k$-cycle $T_j$. If this $k$-cycle
$T_j$ is balanced ($j$ is even) nothing happens. If $T_j$
is imbalanced ($j$ is odd) we change one of its link as
follows: if $j<k$, then $T_j \to \; T_{j-1}$ occurs with
probability $p$, while $T_j \to \; T_{j+1}$ occurs with
probability $1-p$ ; if $j=k$, then $T_{j} \to \; T_{j-1}$ happens
with probability $1$. During one update, the positive [negative]
link which we flip to take a negative [positive] sign is chosen at
random between all the possible positive [negative] links
belonging to the $k$-cycle $T_j$. One unit of time is defined as a
number of updates equal to $L$, where $L$ total number of links of
the network. In Figure \ref{fig:example} we show a simple scheme that
illustrates the dynamical rules in the case $k=4$ (A) and $k=5$
(B). It is evident from the figure that for even values of $k$ the system
remains the same if we simultaneously flip all the spins $s_{i,j}
\to -s_{i,j}$  $\forall \; (i,j)$ and make the transformation $p
\to 1-p$. The same is not true for odd values of $k$. The reason
 is that a $k$-cycle with only ``unfriendly'' links is
balanced for even values of $k$, while it is imbalanced for odd values of $k$. The
presence or absence of this symmetry property for even values of $k$ or odd,
respectively, is responsible for very different features in the
phase structure. This will be studied in detail in the following
sections.

\section{Complete graphs} \label{complete}

We first consider the case of fully connected networks. Later we
extend the main results to the case of diluted networks in
section \ref{sec:diluted}. In a complete graph every individual has a
relationship with everyone else. Let $N$ be the number of nodes of
this complete graph. The total number of  links of the network is
then given by $L={N \choose 2}$, while the total number of
$k$-cycles is given by $M={N \choose k}$. ${x \choose y}$ is the
standard notation of the binomial coefficient. It counts the total
number of different ways of choosing $y$ elements out of $x$
elements in total, while it is $0 \leq y\leq x$ , with $x,y \in \mathbb{N}$. Moreover we
define $M_j$ as the number of $k$-cycles containing $j$ negative
links, and $m_j= \; M_j/\;M$ the respective density of $k$-cycles
of type $T_j$. The total number of positive links $L^+$ is then
related to the number of $k$-cycles by the relation
\begin{equation}
L^+ = \frac{\sum_{i=0}^k\left(k-i\right) \; M_i}{\left(N-2\right)!
\;/\; \left(N-k\right)!}\;\;\; .\label{eq:link_positive}
\end{equation}
A similar relation holds for the total number of negative links
$L^-$
\begin{equation}
L^- = \frac{\sum_{i=0}^k i \; M_i}{\left(N-2\right)! \;/\; \left(N-k\right)!}\;\;\;.
\label{eq:link_negative}
\end{equation}
In particular, in Eq.s (\ref{eq:link_positive}) and
(\ref{eq:link_negative}) the numerators give us the total number
of positive and negative links in all the $k$-cycles,
respectively, while the same denominator comes out from the fact
that one link belongs to
$(N-2)(N-3)\cdots(N-k+1)=\left(N-2\right)!/ \left(N-k\right)!$
different $k$-cycles. Furthermore the density of positive links is
$\rho=L^+/L=1-\sum_{i=0}^k\; i \; m_i$, while the density of
negative links is $1-\rho$.

\subsection{Evolution Equations} \label{evolution}

In view of deriving the mean field equations for the unconstrained
dynamics, introduced in the former section \ref{model}, we need to
define the quantity $M^+_j$ as the average number of $k$-cycles of
type $T_j$ which are attached to a positive link. This number is
given by
\[
M^+_j = \frac{\left(k-j\right) \;\; M_j }{L^+}\;\;\; ,
\]
while similarly
\[
M^-_j = \frac{j \;\; M_j }{L^-}
\]
counts the average number of $k$-cycles of type $T_j$ attached to
a negative link. In term of densities we can easily write
\begin{equation}
m^+_j = \frac{\left(k-j\right) \;\; m_j }{\sum_{i=0}^k\; \left(k-i\right) \; m_i}
\label{eq:density_positive}
\end{equation}
and
\begin{equation}
m^-_j = \frac{j \;\; m_j }{\sum_{i=0}^k\; i \; m_i}\;\;\;.
\label{eq:density_negative}
\end{equation}
Now let $\pi^+$ be the probability that a link flips its sign from
positive to negative in one update event and $\pi^-$ the
probability that a negative link changes its sign to $+1$ in one
update event. We can write such probabilities as
\begin{equation}
\pi^+ = \left(1-p\right) \; \sum_{i=1}^{(k-1)/2} \; m_{2i-1}
\label{eq:prob_positive_odd}
\end{equation}
and
\begin{equation}
\pi^- = p \; \sum_{i=1}^{(k-1)/2} \; m_{2i-1} \; + \; \; m_k \;\;\;,
\label{eq:prob_negative_odd}
\end{equation}
valid for the case odd values of $k$. For even values of $k$, these probabilities read
\begin{equation}
\pi^+ = \left(1-p\right) \; \sum_{i=1}^{k/2} \; m_{2i-1}
\label{eq:prob_positive_even}
\end{equation}
and
\begin{equation}
\pi^- = p \; \sum_{i=1}^{k/2} \; m_{2i-1} \; \; \; .
\label{eq:prob_negative_even}
\end{equation}
Since each update changes $\left(N-2\right)! / \left(N-k\right)!$
$k$-cycles, and also the number of updates in one time step is
equal to $L$ update events, the rate equations in the mean field
approximation can be written as
\begin{equation}
\left\{
\begin{array}{l}
\frac{d}{dt}\; m_0 \; = \; \pi^- \; m^-_1 \; - \;  \pi^+ \; m^+_0
\\
\\
\begin{array}{ll}
\frac{d}{dt}\;m_1 \; = \;  & \pi^+ \; m^+_0 \; + \;  \pi^- \; m^-_2 \; +
\\
  & - \; \pi^- \; m^-_1 \; - \;  \pi^+ \; m^+_1
\end{array}
\\
\vdots
\\
\begin{array}{ll}
\frac{d}{dt}\;m_j \; = \; & \pi^+ \; m^+_{j-1} \; + \;  \pi^- \; m^-_{j+1} \; +
\\
& - \; \pi^- \; m^-_{j} \; - \;  \pi^+ \; m^+_{j}
\end{array}
\\
\vdots
\\
\begin{array}{ll}
\frac{d}{dt}\;m_{k-1} \; = \; & \pi^+ \; m^+_{k-2} \; + \;  \pi^- \; m^-_{k} \; +
\\
& - \; \pi^- \; m^-_{k-1} \; - \;  \pi^+ \; m^+_{k-1}
\end{array}
\\
\\
\frac{d}{dt}\;m_k \; = \; \pi^- \; m^-_{k-1} \; - \;  \pi^- \; m^-_k
\end{array}
\right.
\;\;\;.
\label{eq:mean_field}
\end{equation}
We remark that the only difference between the cases of odd values of $k$ and
even values of $k$ comes from Eq.s (\ref{eq:prob_positive_odd}) and
(\ref{eq:prob_negative_odd}), and Eq.s
(\ref{eq:prob_positive_even}) and (\ref{eq:prob_negative_even}),
respectively. This difference is the main reason why the two cases
odd values of $k$ and even values of $k$ lead to two completely different behavior and
why we treat them separately in the following section
\ref{stationary}.

\subsection{Stationary states} \label{stationary}
Next let us derive the stationary states from the rate equations
(\ref{eq:mean_field}) that give a proper description of the
unconstrained dynamics of $k$-cycles in a complete graph. Imposing
the stationary condition $\frac{d}{dt}\;m_j = 0$ , $\forall \; 0
\leq j \leq k$, we easily obtain
\begin{equation}
m^+_{j-1}\; = \; m^-_j \;\;\;, \; \forall \; 1 \leq j \leq k \;\;\; .
\label{eq:stationary1}
\end{equation}
Then, forming products of the former quantities appearing in Eq.(\ref{eq:stationary1}),
we have
\[
m^+_{j-1}\; m^-_{j+1}\; = \; m^+_j\; m^-_j\;\;\; , \; \forall \; 1 \leq j \leq k \;\;
\]
and, using the definitions of Eq.s (\ref{eq:density_positive}) and
(\ref{eq:density_negative}), we finally obtain
\begin{equation}
 \left(k-j+1\right)\left(j+1\right)
 \; m_{j-1} \; m_{j+1} \; \; = \; \; \left(k-j\right) j \left(\; m_j\; \right)^2\;\;\; ,
\label{eq:stationary3}
\end{equation}
valid $\forall \; 1 \leq j \leq k$. Moreover the normalization
condition  $\sum_i \; m_i \; = \; 1$ should be satisfied.
Furthermore, in the case of  stationary, the density of
friendships should be fixed, so that we should impose that $\pi^+ \; = \;
\pi^-$.


\subsubsection{The case of odd values of $k$}\label{odd}
\begin{figure}[ht]
\includegraphics*[width=0.47\textwidth]{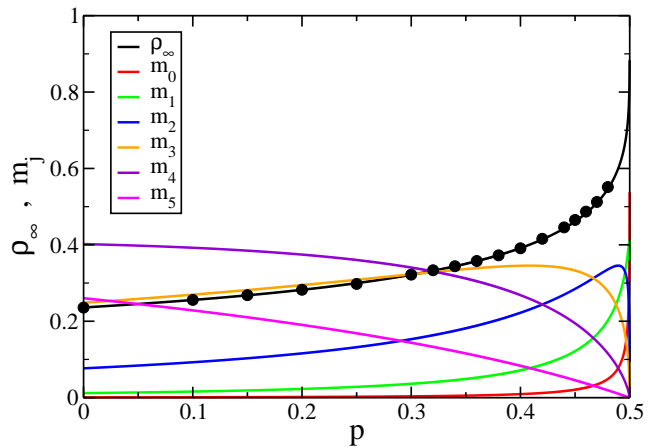}
\caption{(Color online) Exact stationary densities $m_j$ for the
cycles of order $k=5$ from Eq.(\ref{eq:k5a}) and stationary
density of friendly links $\rho_\infty$ from Eq.(\ref{eq:k5b}),
both as a function of the dynamical parameter $p$. Numerical
results are also reported for a system with $N=64$ vertices. Each
value (full dot) is obtained by averaging the density of friendly
links reached after $10^3$ time steps over $10^2$ different
realizations with random initial conditions ($\rho_0 = 0.5$).}
\label{fig:t5}
\end{figure}

\begin{figure}[ht]
\includegraphics*[width=0.47\textwidth]{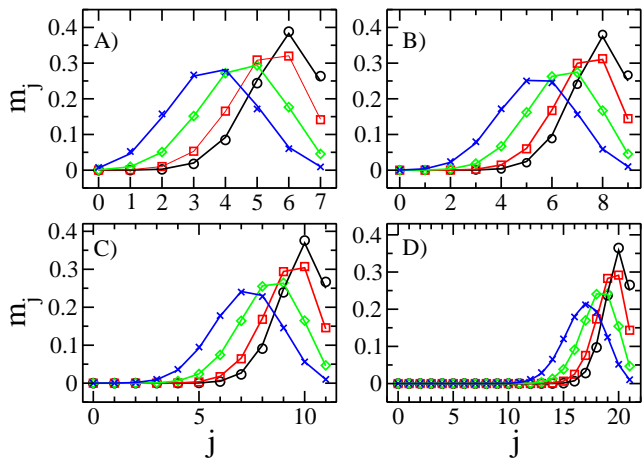}
\caption{(Color online) Stationary densities $m_j$ for the
$k$-cycles with $j$ negative links and different values of $k$  [
$k=7$ (A) , $k=9$ (B), $k=11$ (C) and $k=21$ (D) ], and for
different values of $p$ [ $p=0$ (black circles) , $p=0.3$ (red
squares) , $p=0.44$ (green diamonds) and $p=0.49$ (blue crosses)
]. The numerical results (symbols) represent the histograms
extrapolated from $10^6$ samples and over $50$ different
realizations of the network. In particular the initial values of
the spins are equally likely at each realization (so that
$\rho_0=0.5$), the distributions are sampled after $5\cdot 10^2$
time steps and the system size is always $N=64$. The prediction of
Eq.(\ref{eq:kgen}) is plotted as a full line and the value of
$\rho_\infty$ used is taken from the simulations as the average
value of the stationary density of positive links.}
\label{fig:test}
\end{figure}

\begin{figure}[ht]
\includegraphics*[width=0.47\textwidth]{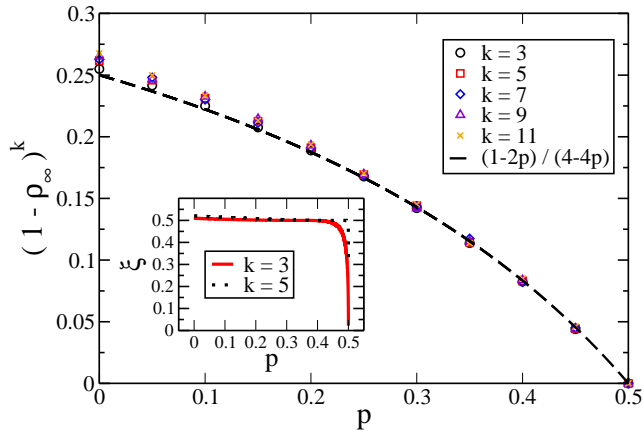}
\caption{(Color online) Numerical results (symbols) and
approximate solution (dashed line) for the function
$\left(1-\rho_\infty\right)^k$, depending on the  stationary
density of positive links $\rho_\infty$ and the parameter $k$ [
$k=3$ (black circles) , $k=5$ (red squares) , $k=7$ (blue
diamonds) , $k=9$ (violet triangles) , $k=11$ (orange crosses) ],
as a function of the dynamical parameter $p$. The theoretical
result, plotted here as a dashed line, is given by
Eq.(\ref{eq:kgen_fin}) for $\xi=1/2$. This prediction is in good
agreement with the numerical results obtained by averaging the
density of friendly links after $10^3$ time steps over $10^2$
different realizations. The system size is $N=64$. Each simulation
starts with random initial conditions ($\rho_0=0.5$). Moreover, as
we can see from the inset, the value of $\xi$ calculated for $k=3$
(red full line) and for $k=5$ (black dotted line) is very close to
$1/2$ for an extended range of $p$.} \label{fig:theory}
\end{figure}

In the case of odd values of $k$, the condition for having a fixed
density of friendships reads
\begin{equation}
m_k \; = \; \left(1 - 2p\right) \; \sum_{i=1}^{(k-1)/2} \; m_{2i-1}\;\;\;,
\label{eq:stationary_odd1}
\end{equation}
where we used Eq.s (\ref{eq:prob_positive_odd}) and
(\ref{eq:prob_negative_odd}). In principle the $k$ equations of
(\ref{eq:stationary3}) plus the normalization condition and the
fixed friendship relation (\ref{eq:stationary_odd1}) determine the
stationary solution. For $k=3$ Antal \emph{et al.} \cite{antal}
found
\begin{equation}
m_j \; = \; {3 \choose j} \; \rho_\infty^{3-j} \; \left(1-\rho_\infty\right)^j\;\;, \; \forall \; 0 \leq j \leq 3 \;\;\; ,
\label{eq:antal1}
\end{equation}
where
\begin{equation}
\rho_\infty \; = \;
\left\{
\begin{array}{ll}
1/\left[ \sqrt{3\left(1-2p\right)} +1 \right] & \textrm{ , if } p \leq 1/2
\\
1 & \textrm{ , if } p \geq 1/2
\end{array}
\right.
\label{eq:antal2}
\end{equation}
is the stationary density of friendly links. In the same manner
also the case $k=5$ can be solved exactly with the solution
\begin{equation}
m_j \; = \; {5 \choose j} \; \rho_\infty^{5-j} \; \left(1-\rho_\infty\right)^j \;\;, \; \forall \; 0 \leq j \leq 5 \;\;\; ,
\label{eq:k5a}
\end{equation}
where
\begin{equation}
\rho_\infty \; = \;
\left[ \sqrt{5\left(1-2p\right) \left(1+\sqrt{1+\frac{1}{5(1-2p)}}\right)} +1 \right]^{-1}
\label{eq:k5b}
\end{equation}
for $p\leq 1/2$, while $\rho_\infty=1$ for $p\geq 1/2$.
\\
In Figure \ref{fig:t5} we plot the densities $m_j$ given by
Eq.(\ref{eq:k5a}) and the stationary density of friendly links
$\rho_\infty$ given by Eq.(\ref{eq:k5b}) as function of $p$.
Moreover we verified the validity of the solution performing
several numerical simulations on a complete graph with $N=64$
nodes (full dots). We compute numerically the average  density of positive
links after $10^3$ time steps, where the average is done over
$10^2$ different realizations of the system. At the beginning of
each realization we select at random the values of the signs of
the links, where each of them has the same probability to be
positive or negative, so that $\rho_0=0.5$. The numerical results
perfectly reproduce our analytical predictions.
\\
As one can easily see, both solutions (\ref{eq:antal1}) and
(\ref{eq:k5a}) are just binomial distributions. This means that
the densities of  a cycle of order $k=3$ or a cycle of order $k=5$
with $j$ negative links are simply given by the probability of
finding these densities on a complete graph in which each link is
set equal to $1$ with probability $\rho_\infty$ or equal to $-1$
with probability $1-\rho_\infty$. (As already noticed in
\cite{antal}, this result may come a bit as a surprise,
because the $3$-cycle or here the $5$-cycle dynamics seems to be
biased towards the reduction of frustration, on the other hand it
is a bias for individual triads without any constraint of the type
that the frustration of the whole "society" should get reduced.)
\\
For odd values of $k>5$, a stationary solution always exists. This
solution becomes harder to find as $k$ increases, because the
maximal order of the polynomials involved increases with $k$ (for $k=3$ we
have polynomials of first order, for $k=5$ polynomials of second
order, for $k=7$ of third and so on). So it becomes impossible to
find the solution analytically as the maximal order of solvable
equations is reached. Nevertheless we can give an approximate
solution using a self-consistent approach as we shall outline in
the following. We suppose that the general solution for the
stationary densities is of the form
\begin{equation}
m_j \; = \; {k \choose j} \; \rho_\infty^{k-j} \; \left(1-\rho_\infty\right)^j \;\;, \; \forall \; 0 \leq j \leq k \;\;\; ,
\label{eq:kgen}
\end{equation}
Eq.(\ref{eq:kgen}) is an appropriate ansatz as we can directly see
from the definition of the density of friendly links $\rho_\infty
= 1-\sum_{i=0}^k i\; m_i= 1- (1-\rho_\infty)$, where the last term
comes out as mean value of the binomial distribution. ( Actually
such self-consistency condition is satisfied by any distribution
of the $m_j$s  with mean value equal to $1-\rho_\infty$. )
Moreover the ansatz for the stationary solution in the form of
Eq.(\ref{eq:kgen}) has the following features: first it is valid
for the special cases $k=3$ and $k=5$, and second, it is
numerically supported. In Figure \ref{fig:test} we show some
results obtained by numerical simulations. We plot 
the densities $m_j$ for different values of $k$ [ $k=7$ (A) ,
$k=9$ (B), $k=11$ (C) and $k=21$ (D) ] and different values of $p$
[ $p=0$ (black circles) , $p=0.3$ (red squares) , $p=0.44$ (green
diamonds) and $p=0.49$ (blue crosses) ]. We performed $50$
different realizations of a system of $N=64$ vertices, where the
densities are extrapolated from  $10^6$ samples ($k$-cycles) at
each realization and after $5\cdot 10^2$ time steps of the
simulations (so that we have reached the stationary state). The
initial values of the signs are chosen to be friendly or
unfriendly with the same probability ($\rho_0=0.5$). The full
lines are given by Eq.(\ref{eq:kgen}) for which the right value of
$\rho_\infty$ is given by the average stationary density of
friendly links and the average is performed over all simulations.
Furthermore, we numerically check whether Eq.(\ref{eq:kgen})
holds, with the same $\rho_\infty$ if we measure the densities of
cycles also of order $k' \neq k$ and moreover, whether it holds
during the time while using the time dependent density of friendly
links $\rho(t)$ instead of the stationary one $\rho_\infty$. Since
all these checks are positive, we may say that if at some time the
distribution of friendly links (and consequently of unfriendly
links) is uncorrelated, it will stay so forever.
\\
Let us assume that the ansatz (\ref{eq:kgen}) is valid, we then
evaluate the unknown value of $\rho_\infty$ self-consistently by
imposing the condition that the density of friendly links is fixed
at the stationary state
\[
\pi^+ \; = \; \pi^- \;\;\; \Leftrightarrow \;\; \left(1-2p\right)
\sum_{i=1}^{(k-1)/2}\; m_{2i-1}\; = \; m_k \;\;\;.
\]
In particular we can write
\begin{equation}
\sum_{i=1}^{(k-1)/2}\;  m_{2i-1} \; + \;  m_k \; = \;
\sum_{i=1}^{(k+1)/2}\;  m_{2i-1} \; = \xi\;, \label{eq:kgen_a}
\end{equation}
and so
\[
m_k\; = \; \left(1-2p\right)\left(\xi-\; m_k\right)\;
\]
from which
\begin{equation}
\rho_\infty\; = \; 1-\left[ \frac{\xi \left(1-2p\right)}{2\left(1-p\right)}\right]^{1/k}\;\;\;,
\label{eq:kgen_fin}
\end{equation}
for $p \leq 1/2$, while $\rho_\infty=1$ for $p \geq 1/2$. In
particular we notice that Eq.(\ref{eq:kgen_fin}) goes to zero as
$k \to \infty$ for $p < 1/2$, because $0 \leq \xi \leq 1$.. This means that in the limit of
large $k$ the stationary density of friendly links takes the
typical shape of a step function centered at $p=1/2$, with
$\rho_\infty=0$ for $p<1/2$ and $\rho_\infty=1$ for $p>1/2$. This
is exactly the result we find for the case even values of $k$ (see the next
section \ref{even}), and it is easily explained since in the limit
of large $k$ the distinction between the cases odd values of $k$ and $k$
even should become irrelevant.
\\
Furthermore it should be noticed that $\xi$ defined in
Eq.(\ref{eq:kgen_a}) is nothing more than a sum of all odd terms
of a binomial distribution. For large values of $k$ we should
expect that the sum of the odd terms is equal to the sum of the
even terms of the distribution, so that
\[
\xi \; = \; \sum_{i=1}^{(k+1)/2}\;
m_{2j-1} \; \simeq \frac{1}{2} \; \simeq \; \sum_{i=0}^{(k-1)/2}\;  m_{2j} \;\;\;,
\]
because of the normalization. In Figure \ref{fig:theory} we plot the
quantity $(1-\rho_\infty)^k$ obtained by numerical simulations for
different values of $k$ [ $k=3$ (black circles) , $k=5$ (red
squares) , $k=7$ (blue diamonds) , $k=9$ (violet triangles),
$k=11$ (orange crosses) ] as a function of $p$. Each point
represents the average value of the density of positive links
(after $10^3$ time steps) over $10^2$ different realizations. The
system size in our simulations is $N=64$, while, at the beginning
of each realization, the links have the same probability to have
positive or negative spin ($\rho_0 = 0.5$). {}From
Eq.(\ref{eq:kgen_fin}) we expect that the numerical results
collapse on the same curve $\xi(1-2p)/(2-2p)$, depending on the
parameter $\xi$. Imposing $\xi=1/2$ [dashed line] we obtain an
excellent fit for all values of $p$. Only for small values of $p$
the fit is less good than for intermediate and large values of
$p$, which is explained by the plot in the inset of
Figure \ref{fig:theory}. There Eq.(\ref{eq:kgen_a}) is shown as
function of $p$ for $k=3$ (black dotted line) and for $k=5$ (red
full line). The values of $m_j$ are taken directly from the
binomial distribution of Eq.(\ref{eq:kgen}) with values of
$\rho_\infty$ known exactly from Eq.s (\ref{eq:antal2}) and
(\ref{eq:k5b}) for $k=3$ and $k=5$, respectively. We can see how
well the approximation $\xi=1/2$ works already for $k=3$ and how
it improves for $k=5$, with the only exception for small values of
$p$ where $\xi > 1/2$. Furthermore we see that $\xi < 1/2$ for $p
\simeq 1/2$, but in this range the dependence on $\xi$ of
Eq.(\ref{eq:kgen_fin}) becomes weaker since the factor $\xi(1-2p)$
tends to zero anyway.

\subsubsection{The case of even values of $k$}\label{even}
\begin{figure}
\includegraphics*[angle=0,width=0.47\textwidth,clip]{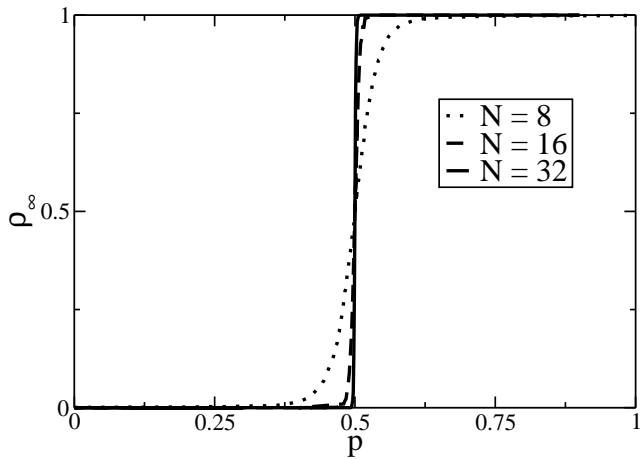}
\caption{Behavior of the stationary density of friendly links
$\rho_{\infty}$ as a function of $p$ for three (small) values of
$N$ [ $N=8$ (dotted line), $16$ (dashed line)  and $32$ (full
line) ] and for $k=4$. The values of the initial configuration are
randomly chosen to be $\pm 1$ with density of friendly links
$\rho_0=0.5$. The curves are obtained from averages over $10^3$
different realizations.} \label{thl}
\end{figure}

The stability of a $k$-cycle with all negative links in the case
of even $k$ (see Figure  \ref{fig:example}) has
deep implications on the global behavior of the model. Actually the
elementary dynamics is now symmetric. Only the value of $p$ gives
a preferential direction (towards a completely friendly or
unfriendly cycle) to the basic processes. With odd $k$, for
$p < 1/2$ the tendency of the dynamics to reach the state with a
minor number of positive links in the elementary processes
(involving no totally unfriendly cycles) is overbalanced by the
process $T_{k}\rightarrow T_{k-1}$ which happens with probability
one, so that in the thermodynamical limit the system ends up in an
active steady state with a finite average density of negative
links due to the competition between the basic processes.
Instead, for even $k$, nothing prevents the system from reaching
the ``hell'', that is a state of only negative links, because here a
completely negative cycle is stable. Only for $p=1/2$ we expect to
find a non-frozen fluctuating final state, since in this case the
elementary dynamical processes are fully symmetric. Imposing the
stationary conditions on the system we do not get detailed
information about the final state. As we can see from Eq.s
(\ref{eq:prob_positive_even}) and (\ref{eq:prob_negative_even}),
for $p\neq1/2$  the only possibility to have $\pi^+=\pi^-$ is the
trivial solution  for which both probabilities are equal to
zero, so that the system must reach a frozen configuration, while
for $p=1/2$, $\pi^+$ and $\pi^-$ are always equal, in this case we
expect the system to reach immediately an active steady state. In
order to describe more precisely the final configuration of this
active steady state, it is instructive to consider the
mean-field equation for the density of positive links. For generic
even value of $k$, it is easy to see that the number of positive
links increases in updates of type $T_{2j-1}\rightarrow
T_{2(j-1)}$ with probability $p$, whereas it decreases in
updates of type $T_{2j-1}\rightarrow T_{2j}$ with probability
$1-p$, so that the mean field equation that governs the behavior
of the density of friendly links is given by
\begin{equation}
\frac{d\rho}{dt}=(2p-1)\rho(1-\rho)\cdot\sum_{i=1}^{k/2}
{k \choose 2i-1}\cdot\rho^{k-2i}(1-\rho)^{2(i-1)}\;\;\;.
\label{mfr}
\end{equation}
For $p\neq1/2$ we have only two stationary states,
$\rho_{\infty}=0$ and $\rho_{\infty}=1$ (the other roots of the
steady state equation are complex). It is easily understood that
for $p<1/2$ the stable configuration is $\rho_{\infty}=0$, while
for $p>1/2$ it is $\rho_{\infty}=1$. In contrast, for $p=1/2$ we
have $\rho(t)=$const at any time, so that
$\rho_{\infty}=\rho(t=0)=\rho_0$. These results are confirmed by
numerical simulations. Moreover, the convergence to the
thermodynamical limit is quite fast, as it can be seen in
Figure  \ref{thl}, where we plot the density of friendly links
$\rho_\infty$ as a function of $p$ for the system sizes $N$ [
$N=8$ (dotted line), $N=16$ (dashed line)  and $N=32$ (full line)
] and for $k=4$. Each curve is obtained from averages over $10^3$
different realizations of the dynamical system. In all simulations
the links get initially assigned the values $\pm 1$ with equal
probability, so that $\rho_0=0.5$.


\subsection{Frozen configurations}\label{frozen}
When all $k$-cycles of the network are balanced we say that the
network itself is balanced. In particular, in the case of our
unconstrained dynamics we can say that if the network is balanced
it has reached a frozen configuration. The configuration is frozen
in the sense that no dynamics is left since the system cannot
escape a balanced configuration. Furthermore it was proven
\cite{cartwright} that if a  graph (not only a complete graph) is balanced it is balanced independently of the choice of $k$ and that the only possible balanced configurations are given by
bipartitions of the network in two subgroups (or ``cliques''),
where all the individuals belonging to the same subgroup are
friends while every couple of individuals belonging to different
subgroups are enemies (this result is also known as
\emph{Structure Theorem} \cite{roberts}). In the case of even values of $k$
the latter result is still valid if all the individuals of one
subgroup are enemies, while two individuals belonging to different
subgroups are friends. It should be noticed that one of the two
cliques may be empty and therefore the configuration of the
paradise (where all the individuals are friends) is also included
in this result, as well as, for the case even values of $k$, the hell with all individuals being
enemies . In the following we will combine our former results about
the stationary states (section \ref{stationary}) with the notion
of frozen configurations in order to predict the probability of
finding a particular balanced configuration and the time needed
for freezing our unconstrained dynamical process. For clarity we
analyze the cases of odd values of $k$ and even values of $k$ separately, again.

\subsubsection{Freezing time for odd values of $k$}
\label{time_odd} 
\begin{figure}
\includegraphics*[width=0.47\textwidth]{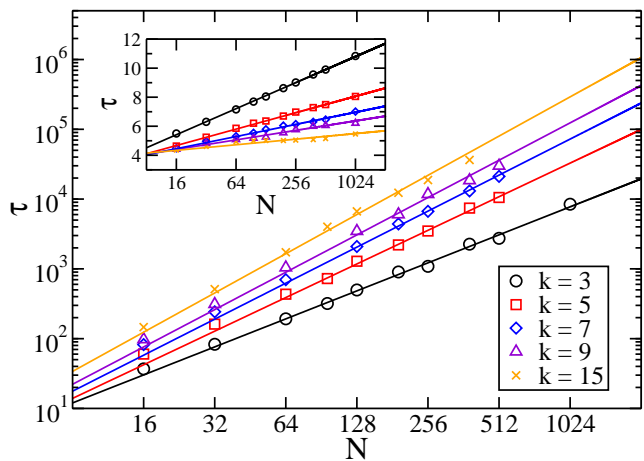}
\caption{(Color online) Numerical results (full dots) for the
freezing time $\tau$ as a function of the system size $N$ and for
various $k$ [ $k=3$ (black circles) , $k=5$ (red squares) , $k=7$
(blue diamonds) , $k=9$ (violet triangles) and $k=15$ (orange
crosses) ]. Each point is given by the average value over several
realizations [ $100$ realizations for sizes $N \leq 64$ , $50$
realizations for $64<N\leq 256$ and $10$ realizations for $N>256$
]. Moreover as initial configuration of each realization the links
are chosen all negative ($\rho_0=0$, antagonistic society) in
order to reduce the statistical error (the standard deviation is
comparable with the symbol size) caused by the small number of
realizations at larger sizes of the system. The full lines have
slope $2(k-1)/k$ as expected from Eq.(\ref{fluc11}). The inset
shows the numerical results for the freezing time $\tau$ , for
different values of $k$ (the same as in the main plot), as a
function of the system size $N$ and for $p=3/4$. Each point of the
inset is given by the average over $10^3$ different realizations
with initial antagonistic society.} \label{fig:time_odd}
\end{figure}

Let $0 \leq N_1 \leq N$ be the size of one of the
two cliques. Therefore the other clique will be of size $N-N_1$.
In such a frozen configuration the total number of positive and
negative links are related to $N_1$ and $N$ by
\begin{equation}
L^+ = \frac{N_1\left(N_1-1\right)}{2}+\frac{\left(N-N_1\right)\left(N-N_1-1\right)}{2}
\label{eq:positive_frozen}
\end{equation}
and
\begin{equation}
L^- = N_1\left(N-N_1\right)
\label{eq:negative_frozen}
\end{equation}
respectively. As we have seen in the former section \ref{odd}, for
odd values of $k$ and $p<1/2$, all the $k$-cycles are uncorrelated during
the unconstrained dynamical evolution, if we start from an
initially uncorrelated configuration. In such cases, we can
consider our system as a purely random process in which the values
of the spins are chosen at random with a certain probability. In
particular,  the probability of a link to be positive is given by
$\rho$ , the density of positive links ( $1-\rho$ is the
probability for a link to be negative). The probability of
reaching a frozen configuration, characterized by two cliques of
$N_1$ nodes and $N-N_1$ nodes, is then given by
\begin{equation}
P(\rho,N_1) = {N \choose N_1}
\rho^{\frac{N(N-1)}{2}-N_1(N-N_1)} \left( 1 -\rho \right)^{N_1(N-N_1)} \;\;\; .
\label{eq:frozen_prob}
\end{equation}
The binomial coefficient ${N \choose N_1}$ in
Eq.(\ref{eq:frozen_prob}) counts the total number of possible
bi-partitions into cliques with $N_1$ and $N-N_1$ nodes ( i.e. the
total number of different ways for choosing $N_1$ nodes out of
$N$), and each of these bi-partitions is considered as equally
likely because of the randomness of the process. We should also
remark that in Eq.(\ref{eq:frozen_prob}) we omit the time
dependence of $\rho$, while the density of positive links $\rho$
follows the following master equation
\[
\frac{d\rho}{dt}=
(1-\rho)^k+(2p-1) \sum_{i=1}^{(k-1)/2} {k \choose 2i-1} \rho^{2i-1} (1-\rho)^{k-2i+1}
\;\;\;.
\]
Eq.(\ref{eq:frozen_prob}) shows that the probability of having a
frozen configuration with cliques of $N_1$ and $N-N_1$ nodes is
extremely small, because the number of the other equiprobable
configurations with the same number of negative and positive links
is equal to ${L \choose L^-} \gg {N \choose N_1}$, where $L^-$
should satisfy Eq.(\ref{eq:negative_frozen}). This allows us to
ignore the transient time to reach the stationary state (we expect
that the system goes to the stationary state exponentially fast
for any $k$, as shown in \cite{antal} for $k=3$) and consider the
probability for obtaining a frozen configurations as
\begin{equation}
P\left(\rho_\infty \right) = \sum_{N_1=0}^N P\left(\rho_\infty,N_1\right) \;\;\; .
\label{eq:frozen_prob_odd}
\end{equation}
This probability provides a good estimate for the order of
magnitude in time $\tau$ that is needed to reach a frozen
configuration, because $\tau \sim 1/P\left(\rho_\infty
\right)$. Unfortunately this estimate reveals that the time needed for
freezing the system becomes very large already for small sizes
$N$ (i.e. $\tau$ increases almost exponentially as a function of $L\sim N^2$). This means that it is practically impossible to verify this
estimate in numerical simulations.
\\
\vspace{5pt}
\\
At the transition, for the dynamical parameter $p=1/2$ we can
follow the same procedure as used  by Antal \emph{et al.}
\cite{antal}. The procedure is based on calculating the time it
takes until a fluctuation in the number of negative links reaches
the same order of magnitude as the average number of negative
links. In this case the systems happens to reach the frozen
configuration of the paradise due to a fluctuation. The number of unfriendly links
$L^- \equiv A(t)$ can be written in the canonical form
\cite{kampen}
\begin{equation}
A(t)=La(t)+\sqrt{L}\eta(t)\;\;\;,
\label{qwe}
\end{equation}
where $a(t)$ is the deterministic part and $\eta(t)$ is a
stochastic variable such that $\langle\eta\rangle=0$. Let us
consider the elementary processes
\begin{equation}
A\longrightarrow\left\{\begin{array}{ll}
    A-1 & \textrm{ , rate } \quad M_k \\
    A-1 & \textrm{ , rate } \quad p\sum_{i=1}^{(k-1)/2}M_{2i-1} \\
    A+1 & \textrm{ , rate } \quad (1-p)\sum_{i=1}^{(k-1)/2}M_{2i-1}
\end{array}
\right.
\label{elpr1}
\end{equation}
and therefore
\begin{equation}
A^2\longrightarrow\left\{\begin{array}{ll}
    A^2-2A+1 & \textrm{ , rate} \quad  M_k \\
    A^2-2A+1 & \textrm{ , rate} \quad p\sum_{i=1}^{(k-1)/2}M_{2i-1} \\
    A^2+2A+1 & \textrm{ , rate} \quad (1-p)\sum_{i=1}^{(k-1)/2}M_{2i-1}
\end{array}
\right.
\;.
\label{elpr2}
\end{equation}
We can then write the following equations for the mean values of
$A$ and $A^2$
\[
\frac{d\langle A\rangle}{dt}=-\langle M_k\rangle+(1-2p)\sum_{i=1}^{(k-1)/2}\langle M_{2i-1}\rangle
\]
and
\[
\begin{array}{ll}
\frac{d\langle A^2\rangle}{dt}= & \langle(1-2A)M_k\rangle+
\\
& +p\left\langle(1-2A)\sum_{i=1}^{(k-1)/2}M_{2i-1}\right\rangle+
\\
& +(1-2p)\left\langle(1+2A)\sum_{i=1}^{(k+1)/2}M_{2i-1}\right\rangle
\end{array}
\;\;\;.
\]
For $p=1/2$ we obtain
\begin{equation}
\frac{d\langle A\rangle}{dt}=-\langle M_k\rangle
\label{fluc3}
\end{equation}
and
\[
\frac{d\langle A^2\rangle}{dt}=\langle M_k\rangle+\sum_{i=1}^{(k-1)/2}
\langle M_{2i-1}\rangle-2\langle AM_k\rangle \;\;\;.
\]
Since it is $\langle A\rangle\sim a$ and $\langle M_k\rangle\sim
a^k$, we get from Eq.(\ref{fluc3})
\begin{equation}
\frac{da}{dt}=-a^k \;\;\; ,
\label{fluc4B}
\end{equation}
from which
\begin{equation}
a(t)\sim t^{-\frac{1}{k-1}}\;\;\;.
\label{fluc5}
\end{equation}
On the other hand, considering that $d\langle A\rangle^2/dt=2\langle
A\rangle\cdot d\langle A\rangle/dt$ and by definition
$\sigma=\langle A^2\rangle-\langle A\rangle^2=\langle\eta^2\rangle$, we have
\begin{equation}
\frac{d\sigma}{dt}=\langle M_k\rangle+\sum_{i=1}^{(k-1)/2}
\langle M_{2i-1}\rangle-2(\langle AM_k\rangle-\langle A\rangle\langle M_k\rangle)\;\;\;.
\label{fluc6}
\end{equation}
Moreover we can write
\[
\begin{array}{rl}
\langle AM_k\rangle-\langle A\rangle\langle M_k\rangle=
&\langle(La+\sqrt{L}\eta)M_k\rangle-La\langle M_k\rangle=
\\
= &  \sqrt{L}\langle\eta M_k\rangle
\end{array}\;\;\;.
\]
It is easy to see that $\langle\eta M_k\rangle\sim\langle
\eta A^k\rangle=\langle\eta(La+\sqrt{L}\eta)^k\rangle$,
so that
\begin{equation}
\begin{array}{ll}
\langle\eta M_k\rangle\sim & \langle\eta\cdot(L^ka^k+kL^{k-1/2}a^{k-1}
\eta+\dots+L^{k/2}\eta^k)\rangle=
\\
& =  kL^{k-1/2}a^{k-1}\langle\eta^2\rangle+O(\langle\eta^3\rangle)\;\;\;.
\end{array}
\label{fluc8}
\end{equation}
Dividing Eq.(\ref{fluc6}) by Eq.(\ref{fluc4B}) and using Eq.(\ref{fluc8}) we get
\begin{equation}
\frac{d\sigma}{da}=-\left[2ka^{k-1}\sigma-\sum_{i=1}^{(k+1)/2}
{k\choose 2i-1}a^{2i-1}(1-a)^{k-2i+1}\right]. \label{fluc9}
\end{equation}
Here we have taken into account that
\begin{equation}
\langle M_j\rangle\sim{k\choose j}a^j(1-a)^{k-j}\;\;\;.
\label{eq:a}
\end{equation}
It is straightforward to find the solution of Eq.(\ref{fluc9}) as
\[
\sigma(a)=Ca^{2k}+\frac{\gamma_k}{a}+\dots+\frac{\gamma_0}{a^{k-2}}\;\;\;,
\]
with $C$ and $\gamma_j$ suitable constants. From  Eq.(\ref{fluc5}), for $t\to \infty$ we have
\[
\sigma\sim a^{-(k-2)}\sim t^{\frac{k-2}{k-1}}\;\;\; .
\]
For $\eta\sim\sqrt{\sigma}$, we finally obtain
\[
\eta\sim t^{\frac{k-2}{2(k-1)}}\;\;\;.
\]
In general, the system will reach the frozen state of the paradise when the
fluctuations of the number of negative links become of the same
order as its mean value. (Note that in this case the mean-field
approach is no longer valid.) Then, in order of finding the
freezing time $\tau$ we have just to set equal the two terms on
the right hand side of Eq.(\ref{qwe}).
\begin{equation}
La(\tau)\sim\sqrt{L}\eta(\tau)\;\;\;.
\label{eq:condition}
\end{equation}
Since $L\sim N^2$, we get a power-law behavior
\begin{equation}
\tau\sim N^\beta
\label{eq:freezing_time_odd}
\end{equation}
with exponent $\beta$ as a function of $k$ according to
\begin{equation}
\beta=2\frac{k-1}{k}\;\;\;.
\label{fluc11}
\end{equation}
It is worth noticing that in the limit $k \to \infty$ we obtain
$\beta=2$, which is the same result as in the case of even values of $k$ as
we shall see soon. The analytical results of this subsection are
well confirmed by simulations, cf. Figure \ref{fig:time_odd}. There
we study numerically the freezing time $\tau$ as a function of the
system size $N$ for different odd values of $k$  [ $k=3$ (black
circles) , $k=5$ (red squares) , $k=7$ (blue diamonds) , $k=9$
(violet triangles) and $k=15$ (orange crosses) ]. The freezing
time is measured until all links have
positive sign and paradise is reached. Other frozen
configurations are too unlikely to be realized. Each point  stands for the average value over a different number of
realizations of the dynamical system [ $100$ realizations for
sizes $N \leq 64$ , $50$ realizations for $64<N\leq 256$ and $10$
realizations for $N>256$ ], where the initial configuration is
always chosen as an antagonistic society (all the links being
negative so that $\rho_0=0$)  to
reduce the statistical error. The standard deviations around
the averages have sizes comparable with the symbol sizes. The full
lines stands for power laws with exponents given by
Eq.(\ref{fluc11}). They perfectly fit with the numerical
measurements.

\vspace{1cm}

For $p>1/2$ the freezing time $\tau$ scales as
\begin{equation}
\tau \sim \ln{N}\;\;\;.
\label{eq:time_pbigger}
\end{equation}
The derivation would be the same as in the paper of Antal \emph{et
al.} \cite{antal}. It should be noticed that for $p>1/2$ the
paradise is reached as soon as $k$ increases. For simplicity let
$p=1$ and imagine that the system is at the closest configuration
to the paradise, for which only one link in the system has
negative spin. This link belongs to $R = (N-2)!/(N-k)!$ different
$k$-cycles. At each update event we select one $k$-cycle at random
out of $M = {N \choose k}$ total $k$-cycles. This way we have to wait a
number of update events $E \sim M/R$ until the paradise is reached, which leads to a freezing
time $\tau \sim E/L$, with $L$ the total number of links
independent on $k$, so that
\begin{equation}
\tau \sim \frac{1}{k!}\;\;\;.
\label{eq:time_biggerp}
\end{equation}
For values of $1/2<p<1$ the $k$-dependence of $\tau$ should be
weaker than the one in Eq.(\ref{eq:time_biggerp}), but anyway
$\tau$ should be a decreasing function of $k$. The inset of
Figure \ref{fig:time_odd} shows the numerical results obtained for
$p=3/4$ as a function of the size of the system $N$. The freezing
time $\tau$ is measured for different values of $k$. We plot the average value  over $10^3$ different
realizations with initial condition $\rho_0=0$.

\subsubsection{Freezing time for even values of $k$}
\label{sec:time_even}
\begin{figure}
\includegraphics*[width=0.47\textwidth]{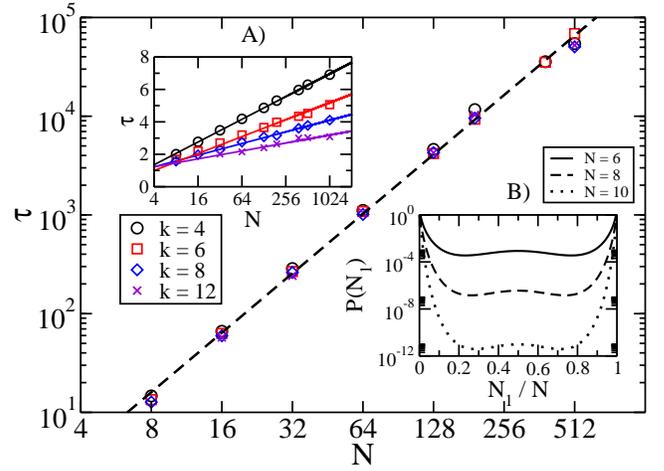}
\caption{(Color online) Numerical results  for the
freezing time $\tau$ as a function of the system size $N$ and for
various even values of $k$ [ $k=4$ (black circles) , $k=6$ (red
squares) , $k=8$ (blue diamonds) and $k=12$ (violet crosses) ] and
for $p=1/2$. Each point is given by the average value over several
realizations [ $100$ realizations for sizes $N \leq 64$ , $50$
realizations for $64<N\leq 256$ and $10$ realizations for $N>256$
]. Moreover, at the beginning of each realization the links are
chosen to be positive or negative with the same probability
($\rho_0=0.5$). The dashed line has, in the log-log plane, slope
$2$ as expected in Eq.(\ref{eq:freezing_time_even}). The inset A)
shows the numerical results for the freezing time $\tau$ , for
different values of $k$ (the same as in the main plot), as a
function of the system size $N$ and for $p=3/4$. Each point of the
inset is given by the average over $10^3$ different realizations
with random initial conditions. The full lines are all
proportional to $\ln{N}$ as expected. The inset B) shows the
not-normalized probability $P(N_1)$ as a function of the ratio
$N_1/N$ and for different values of the system size $N$ [ $N=6$
(full line), $N=8$ (dashed line) and $N=10$ (dotted line) ]. As
one can see, $P(N_1)$ is extremely small for values of $0 < N1 <
N$ already for $N=10$.} \label{fig:time_even}
\end{figure}

In the case of even values of $k$ and $p=1/2$ the master equation for the
density of positive links [ Eq.(\ref{mfr}) ] reads as
$d\rho/dt=0$. Therefore, the density of friendly links, $\rho$,
should be constant during time for an infinite large system. In
finite size systems the dynamics is subjected to non-negligible
fluctuations. This allows to understand the scaling features of the freezing time $\tau$ with the system size. The order of the fluctuations is $\sqrt{L}$ because the process is completely random as we have seen for the case odd values of $k$ and $p<1/2$. Differently from the latter case, for even values of $k$ and $p=1/2$ the system has  no tendency to go to a fixed point determined by $p$ because $d\rho/dt=0$.  We can view the
dynamical system as a Markov chain, with discrete steps in time
and state space, for which the transition probability for passing
from a state with $L^-(t-1)$ negative at time $t-1$ to a state
with $L^-(t)$ negative links at time $t$ is given by
\begin{equation}
\begin{array}{l}
P\left[ \; L^-(t)\; | \; L^-(t-1) \; \right] =
\\
=  {L \choose L^-(t)} \left(\frac{L-L^-(t-1)}{L} \right)^{L-L^-(t)} \left(\frac{L^-(t-1)}{L} \right)^{L^-(t)}\;\;\;.
\end{array}
\label{eq:markov}
\end{equation}
So that the probability of having $L^-(t)$ negative links at time $t$ is just a binomial distribution where the probability of having one negative link is given by $\frac{L^-(t-1)}{L}$, the density of negative links at time $t-1$. This includes both the randomness of the displacement of negative links and the absence of a particular fixed point dependent on $p$. The Markov process, with transition probability given by Eq.(\ref{eq:markov}), is known under the name of the Wright-Fisher
model \cite{wright} from the context of biology. The Wright-Fisher
model is a simple stochastic model for the reproduction of diploid
organisms (diploid means that each organism has two genes, here
named as ``$-$'' and ``$+$''), it was proposed independently by R.A. Fisher and S. Wright at the
beginning of the thirties \cite{wright}. The population size of
genes in an organism is fixed and equal to $L/2$ so that the total
number of genes is $L$. Each organism lives only for one
generation and dies after the offsprings are made. Each offspring
receives two genes, each one selected with probability
$1/2$ out of the two genes of a parent of which two are
randomly selected from the population of the former generation.
Now let us assume that there is a random initial configuration of
positive and negative genes with a slight surplus of negative
genes. The offspring generation selects its genes randomly from
this pool and provides the pool for the next offspring generation.
Since the pools get never refreshed by a new random configuration,
the initial surplus of negative links gets amplified in each
offspring generation until the whole population of genes is
"negative". Actually the solution of the Wright-Fisher model is
quite simple. The process always converges to a final state with
$L^-=0$ [$L^+=L$] or $L^-=L$ [$L^+=0$], corresponding to our
heaven and [hell] solutions for even values of $k$. The average value over
several realizations of the same process depends on the initial
density of friendly links $\rho_0$ according to
\[
\langle L^- \rangle = \rho_0 \delta(0) + (1-\rho_0) \delta(L)\;\;\; ,
\]
where $\delta(x)=1$ for $x=0$ and $\delta(x)=0$ otherwise.
Furthermore, on average, the number of negative links decays
exponentially fast to one of the two extremal values
\[
\langle L^- (t) \rangle \simeq L
\left\{
\begin{array}{l}
e^{-t/L}
\\
1 - e^{-t/L}
\end{array}
\right.   \;\;\;.
\]
with typical decay time
\begin{equation}
\tau \sim L \sim N^2\;\;\;.
\label{eq:freezing_time_even}
\end{equation}
This result is perfectly reproduced by the numerical data plotted
in Figure \ref{fig:time_even}. The main plot shows the average time
needed to reach a balanced configuration as a function of the size
of the system $N$ and for different values of $k$ [ $k=4$ (black
circles) , $k=6$ (red squares) , $k=8$ (blue diamonds) and $k=12$
(violet crosses) ]. The averages are performed over different
numbers of realizations depending on the size $N$ [ $1000$
realizations for sizes $N \leq 128$ , $500$ realizations for
$128<N\leq 384$ and $50$ realizations for $N=384$ and $N=512$, and
$10$ realizations for $N=1024$ ]. The dashed line in
Figure \ref{fig:time_even} has, in the log-log plane, a slope equal
to $2$, all numerical data fit very well with this line.
Furthermore it should be noticed that there is no $k$-dependence
of the freezing time $\tau$, as it is described by
Eq.(\ref{eq:markov}). This is reflected by the fact that $\tau$ is
the same for all the values of $k$ considered in the numerical
measurements.
Nevertheless there is a difference between our model and the
Wright-Fisher model that should be noticed. During the evolution
of our model there is the possibility that the system  freezes in
a configuration different from the paradise ( $L^-=0$ ) or the
hell ($L^-=L$ ). The probability of this event is still given by
Eq.(\ref{eq:frozen_prob}), with $r=L^+(N_1)/L$ as the stationary
condition [ $L^+(N_1)$ is given by Eq.(\ref{eq:positive_frozen})
]. In this way Eq.(\ref{eq:frozen_prob}) gives us $P(N_1)$,
the not-normalized probability for the system to freeze in a
balanced configuration with two cliques of $N_1$ and $N-N_1$
nodes, respectively. It is straightforward to see that $P(N_1)=1$
for $N_1=0$ or for $N_1=N$, so that the paradise has a
non-vanishing probability to be a frozen configuration.
Differently for any other value of $0 < N_1 < N$, $P(N_1)$
decreases to zero faster than $1/N$. This means that for values of
$N$ large enough it is appropriate to forget about the
intermediate frozen configurations and to consider the features of
our model as being very well approximated by those of the
Wright-Fisher model. In the inset B) of Figure \ref{fig:time_even}
the function $P(N_1)$ is plotted for different values of $N$ [
$N=6$ (full line), $N=8$ (dashed line) and $N=10$ (dotted line) ]
with $N_1$ a continuous variable for clarity of the figure (we
approximate the factorial with the Stirling's formula). Obviously
$P(N_1)$ disappears for $0 < N_1 < N$ as $N$ increases, already
for reasonably small values of $N$.
\\
The dependence $\tau \sim N^2$  can also be obtained using the
same procedure as the one in section \ref{time_odd} for the case
odd values of $k$ and $p=1/2$. In particular for even values of $k$ we can rewrite
Eq.(\ref{elpr1}) according to \begin{equation}
A\longrightarrow\left\{\begin{array}{ll}
    A-1 & \textrm{ , rate}\quad p\sum_{i=1}^{k/2}M_{2i-1} \\
    A+1 & \textrm{ , rate}\quad (1-p)\sum_{i=1}^{k/2}M_{2i-1}
\end{array}
\right .
\label{elpr1_even}
\end{equation}
and therefore Eq.(\ref{elpr2}) according to
\begin{equation}
A^2\longrightarrow\left\{\begin{array}{ll}
    A^2-2A+1 & \textrm{ , rate}\quad p\sum_{i=1}^{k/2}M_{2i-1} \\
    A^2+2A+1 & \textrm{ , rate}\quad (1-p)\sum_{i=1}^{k/2}M_{2i-1}
\end{array}
\right. \;\;\;.
\label{elpr2_even}
\end{equation}
For $p=1/2$ we have
\begin{equation}
\frac{d\langle A\rangle}{dt}=0
\label{fluq1_even}
\end{equation}
and
\[
\frac{d\langle A^2\rangle}{dt}=\sum_{i=1}^{k/2} \langle M_{2i-1} \rangle\;\;\;.
\]
Eq.(\ref{fluq1_even}) tells us that $a\sim \langle A\rangle=$const, so that we have
\[
\eta\sim \sqrt{t}\;\;\;,
\]
remebering Eq.(\ref{eq:a}). As in the previous case, for  determining the freezing time we
impose the condition that the average value is of the same order
as the fluctuations [Eq.(\ref{eq:condition})], and, for $L\sim
N^2$, we obtain again Eq.(\ref{eq:freezing_time_even}).
\\
\vspace{5pt}
\\
For even values of $k$ and for $p \neq 1/2$ the time $\tau$ needed
for reaching a frozen configuration scales as $\tau \sim \ln{N}$.
In the inset of Figure \ref{fig:time_even} numerical estimates of
$\tau$ for $p=3/4$ and different values of $k$ demonstrate this
dependence on the size $N$ of the system. Each point  is
obtained from averaging over $10^3$ different simulations with the
same initial conditions $\rho_0=0.5$. Again, as in the case of $k$
odd and $p>1/2$, $\tau$ is a decreasing function of $k$ and the
same argument used for obtaining Eq.(\ref{eq:time_biggerp}) can be
applied here.

\section{Diluted Networks}\label{sec:diluted}
In this section we extend the former results, valid in the case of
fully connected networks, to diluted networks. Real networks,
apart from very small ones, cannot be represented by complete
graphs. The situation in which all individuals know each other is
in practice very unlikely. As mentioned in the introduction,
links may be also missing, because individuals neither like nor
dislike each other but are just indifferent. In the following we
analyze the features of dynamical systems, still following the
unconstrained $k$-cycle dynamics, but living on topologies given
by diluted networks.
\\
For diluted networks there is an interesting connection to another
set of problems that leads to a new interpretation of the social
balance problem in terms of a certain $k$-SAT ($k$S) problems (SAT stands
for satisfiability) \cite{cook,mezard,mezard2}. In such a problem a formula $F$ consists of $Q$
logical clauses $\left\{C_q\right\}_{q=1,\ldots,Q}$ which are
defined over a set of $B$ Boolean variables
$\left\{x_i=0,1\right\}_{i=1,\ldots,B}$ which can take two
possible values $0=$\emph{FALSE} or $1=$\emph{TRUE}. Every clause
contains $k$ randomly chosen Boolean variables that are connected
by logical $OR$ operations ($\bigvee$). They appear negated with a
certain probability. In the formula $F$, all clauses are connected
by logical $AND$ operations ($\bigwedge$)
\[
F=\bigwedge_{q=1}^Q C_q\;\;\;,
\]
so that all clauses $C_q$ should be simultaneously satisfied in
order to satisfy the formula $F$. A particular formulation of the
$k$S problem is the $k$-XOR-SAT ($k$XS) problem \cite{weigt,semerjian,mezard2,cocco}, in which each clause
$C_q$ is a parity check of the kind
\begin{equation}
C_q= x_{i_1}^q+x_{i_2}^q+ \ldots + x_{i_k}^q \;\;\; \textrm{mod }2\;\;\;,
\label{eq:xor}
\end{equation}
so that $C_q$ is \emph{TRUE} if the total number of true variables
which define the clause is odd, while otherwise the clause $C_q$
is \emph{FALSE}. It is straightforward to map the $k$XS
problem to our former model for the case odd values of $k$. Actually, each
clause $C_q$ corresponds to a $k$-cycle [$Q \equiv M$] and each
variable $x_v$ to a link $(i,j)$. Furthermore [$B \equiv L$] with
the correspondence $s_{i,j}=1$ for $x_v=1$, while $s_{i,j}=-1$ for
$x_v=0$. For the case of even values of $k$, one can use the same mapping
but consider as clause $C_q$ in Eq.(\ref{eq:xor}) its negation
$\overline{C_q}$. In this way, when the number of satisfied
variables $x_i^q$ is odd the clause $\overline{C_q}$ is
unsatisfied for odd values of $k$, while $\overline{C_q}$ is satisfied for
even values of $k$.
\\
Moreover a typical algorithm for finding a solution of the $k$S
problem  is the so-called Random-Walk SAT (RWS). The procedure is
the following \cite{weigt,semerjian}: select one unsatisfied clause $C_q$
randomly, next invert one randomly chosen variable of its $k$
variables $x_{i^*}^q$; repeat this procedure until no unsatisfied
clauses are left in the problem. Each update is counted as $1/B$
units of time. As one can easily see, this algorithm is very
similar to our unconstrained dynamics apart from two aspects.
First, in our unconstrained dynamics we use the dynamical
propensity parameter $p$, while it is absent in the RWS. Second,
in our unconstrained dynamics we count also the choice of a
balanced $k$-cycle as update event, although it does not change
the system at all. Because of this reason, the literal application of
the original algorithm of unconstrained dynamics has very high
computational costs if it is applied to diluted networks. Apart
from the parameter $p$, we can therefore use the same RWS
algorithm for our unconstrained dynamics of $k$-cycles. This
algorithm is more reasonable because it selects at each update
event only imbalanced $k$-cycles which are actually the only ones
that should be updated. In case of an all-to-all topology there
are so many triads that a preordering according to the property of
being balanced or not is too time consuming so that in this case
our former version is more appropriate. In order to count the
time as in our original framework of the unconstrained dynamics,
we should impose that, at the $n$-th update event, the time
increases as
\begin{equation}
t_n\; \; = \;\; t_{n-1}\;\;+ \;\;\frac{1}{L}\; \cdot \; \frac{\alpha}{\alpha^{(n-1)}_{u}}\;\;\;.
\label{eq:time_scaling}
\end{equation}
Here $\alpha=M/L$ stands for the ratio of the total number of
$k$-cycles of the system (i.e. total number of clauses) and the
total number of links (i.e. total number of variables). The
parameter $\alpha$ is called the ``dilution'' parameter, it can
take all possible values in the interval $\left[0 , {L \choose
k}/L\right]$. $\alpha^{(n-1)}_{u}=\sum_{i=1}^{(k+1)/2}M_{2i-1}/L$
is the ratio of the total number of imbalanced (or
``unsatisfied'') $k$-cycles over the total number of links, in
particular $\alpha^{(n-1)}_{u}$ is computed before an instant of
time at which the $n$-th update event is implemented. Therefore the
ratio $\alpha/\alpha^{(n-1)}_{u}$ gives us the inverse of the
probability for finding an imbalanced $k$-cycle, out of all, balanced or imbalanced, $k$-cycles, at the $n$-th
update event. This is a good approximation to the time defined in
the original unconstrained dynamics. It should be noticed that
this algorithm works faster in units of this computational time,
but the simulation  time should be counted in the same units as defined for the unconstrained
dynamics introduced in section \ref{model}.
\\
The usual performance of the RWS is fully determined by the
dilution parameter $\alpha$. For $\alpha \leq \alpha_d$ the RWS
always finds a solution of the $k$S problem within a time that
scales linearly with the number of variables $L$. In particular
for the $k$XS problem $\alpha_d=1/k$. For $\alpha_d < \alpha <
\alpha_c$ the RWS is still able to find a solution for the $k$S
problem, but the time needed to find the solution grows
exponentially with the number of variables $L$. For the case of
the $3$XS problem $\alpha_c \simeq 0.918$. $\alpha_d$ is the value
of the dilution parameter for which we have the ``dynamical''
transition, depending on the dynamics of the algorithm while
$\alpha_c$ represents the transition between the SAT and the UNSAT
regions: for values of $\alpha \geq \alpha_c$ the RWS is no longer
able to find any solution for the $k$S problem, and in fact no
such solution with zero frustration exists for the $k$S
problem. Furthermore there is a third critical threshold
$\alpha_s$, with $\alpha_d < \alpha_s <\alpha_c$. For values of
$\alpha < \alpha_s$ all solutions of the $k$S problem found by
the RWS are located into a large cluster of solutions and the
averaged and normalized Hamming distance inside this cluster is
$\langle d \rangle \simeq 1/2$. For $\alpha > \alpha_s$ the
solutions space splits into a number of small clusters (that grows
exponentially with the number of variables $L$) , for which the
averaged and normalized Hamming distance inside each cluster is
$\langle d \rangle \simeq 0.14$, while  the averaged and
normalized Hamming distance between two solutions lying in
different clusters is still $\langle d \rangle \simeq 1/2$ \cite{cocco}. For
the special case of the $3$XS problem $\alpha_s$ was found
as $\alpha_s\simeq 0.818$.
\\
In order to connect the problems of social balance on diluted
networks and the $k$XS problem on a diluted system we shall first
translate the parameters into each other. First of all we need to
calculate the ratio $\alpha=M/L$ between the total number of
$k$-cycles of the network and the total number of links $L$
(section \ref{sec:ratio}). Next we consider the standard RWS
applied to the $k$XS problem taking care of the right way
of computing the time as it is given by the rule
(\ref{eq:time_scaling}) and the introduction of the dynamical
parameter $p$ (section \ref{sec:RWS}). In particular we focus on
the ``dynamical'' transition at $\alpha_d$ (section
\ref{sec:alphad}) and the transition in solution space concerning
the clustering properties of the solutions at $\alpha_s$ (section
\ref{sec:alphas}). The dynamical parameter $p$, formerly called
the propensity parameter, leads to a critical value $p_c$ above
which it is always possible to find a solution  within a time that grows at most linearly with the
system size (section
\ref{sec:pc}). Finally, in section \ref{sec:alpham} we decrease
the dilution, i.e. increase $\alpha$ to $\alpha_m$ such that for
$\alpha \geq \alpha_m$ the system is fully described by the mean
field equations of the former sections. We focus on the simplest
case $k=3$, but all results presented here for $k=3$ should be
qualitatively valid for any value of $k \geq 3$.

\subsection{Ratio $\alpha$ for random networks}
\label{sec:ratio} 
\begin{figure}
\includegraphics*[width=0.47\textwidth]{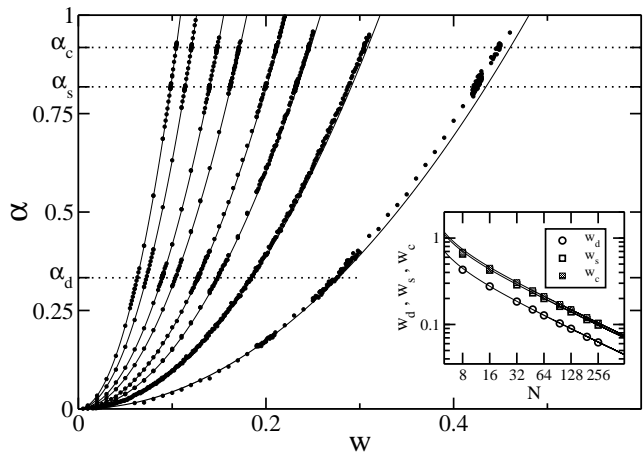}
\caption{Numerical results (full dots) for the ratio $\alpha=M/L$
between the total number of cycles $M$ of order $k=3$ and the
total number of links $L$ as a function of the probability $w$ for
different sizes of Erd\"os-R\'enyi networks. In particular the
numerical results refer to different network size $N$: from bottom
to top $N=16, 32, 48, 64, 96, 128, 192$ and $256$. Each point is
given by the average over $10^3$ network realizations. The full
lines are the predicted values given by
Eq.(\ref{eq:average_ratio}), while the dotted lines denote the
critical values $\alpha_d =1/3$ , $\alpha_s =0.818$ and $\alpha_c
=0.918$ as described in detail in the text. In particular the
numerical values of the probability $w$ for which these three
critical values of $\alpha$ are realized are denoted by $w_d$
(open circles), $w_s$ (open squares) and $w_c$ (gray squares)
respectively, they are plotted in the inset, where the full lines
are extrapolated by Eq.(\ref{eq:average_ratio}) as $w_i =
\sqrt{3\alpha_i/N} \; , \; i=d,s,c$. The two upper curves for
$w_s$ and $w_c$ almost coincide.} \label{fig:ratio}
\end{figure}

Let us first consider Erd\"os-R\'enyi networks \cite{erdos} as
a diluted version of the all-to-all topology that we studied so
far. An Erd\"os-R\'enyi network, or a random network, is a network
in which each of the ${N \choose 2}$ different pairs of nodes is
connected with probability $w$. The average number of links is
simply $\langle L \rangle = w {N \choose 2}$. The average number
of cycles of order $k$ is given $\langle M \rangle = w^k {N
\choose k}$, so that the average ratio $\langle \alpha \rangle$
can be estimated as
\begin{equation}
\langle \alpha \rangle \simeq w^{k-1} \frac{2N^{k-2}}{k!} \;\;\; .
\label{eq:average_ratio}
\end{equation}
In Figure  \ref{fig:ratio} we plot the numerical results obtained
for the ratio $\alpha$ as a function of the probability $w$, in
the particular case of cycles of order $k=3$. The reported
results, from bottom to top, have been obtained for values of
$N=16, 32, 48, 64, 96, 128, 192$ and $256$. Each point is given by
the average over $10^3$ different network realizations. In
particular these numerical results fit very well with the
expectations (full lines) of Eq.(\ref{eq:average_ratio}),
especially for large values of $N$ and/or small values of $w$.
Furthermore the critical values $\alpha_d =1/3$ , $\alpha_s
=0.818$ and $\alpha_c =0.918$ (dotted lines)  are used for
extrapolating the numerical results of $w_d$ (open circles), $w_s$
(open squares) and $w_c$ (gray squares) respectively [see the
inset of Figure  \ref{fig:ratio}]. $w_i \; , \; i=d,s,c$ is the
value of the probability for which the ratio $\alpha_i \; , \;
i=d,s,c$ is satisfied. As expected, they follow the rule $w_i =
\sqrt{3\alpha_i/N} \; , \; i=d,s,c$ predicted by
Eq.(\ref{eq:average_ratio}) for $k=3$.
\\
According to the isomorphism traced between the $k$XS problem and the social balance for $k$-cycles, from now on we will not make any distinction between the words problem and network, variable and link, $k$-clause and $k$-cycle, value and sign (or spin), false and negative (or unfriendly), true and positive (or friendly), satisfied and balanced (or unfrustrated), unsatisfied and imbalanced (or frustrated), etc\ldots . 

\subsection{$p$-Random-Walk SAT }
\label{sec:RWS}
So far we have established the connection between
the $k$XS problem and the  social
balance for $k$-cycles, proposed in this paper. In particular we have determined
how the dilution parameter $\alpha$ is related to diluted random
networks parameterized by $w$. In this section we extend the known
results for the standard RWS of \cite{weigt,semerjian} to the $p$-Random-Walk SAT ($p$RWS) algorithm, that is the RWS algorithm extended by the dynamical parameter $p$ that played the role of a propensity parameter in connection with the social balance problem. The steps of the $p$RWS are as follows:
\begin{enumerate}
\item{Select randomly a frustrated clause between all frustrated clauses.}
\item{Instead of randomly inverting the value of  one of its $k$ variables, as for an update in the case of the RWS, apply the following procedure:
\begin{itemize}
\item{if the clause contains both true and false variables, select with probability $p$ one of its false variable, randomly chosen between all the false variables belonging to the clause, and flip it to the true value;}
\item{if the clause contains both true and false variables, select with probability $1-p$ one of its true variable, randomly chosen between all the true variables belonging to the clause, and flip it to the false value;}
\item{if the clause contains only false values ($k$ should be odd), select with probability $1$ one of its false variables, randomly chosen between all the false variables belonging to the clause, and flip it to the true value.}
\end{itemize}}
\item{Go back to point 1 until no unsatisfied clauses are present in the problem.}
\end{enumerate}
The update rules of point 2 are the same used in the case of $k$-cycle dynamics and illustrated in Figure  \ref{fig:example} for the cases $k=4$ (A) and $k=5$ (B). For the special case of $3$XS problem, the standard RWS algorithm and the $p$RWS algorithm coincides for the dynamical parameter $p=1/3$.

\subsubsection{Dynamical transition at $\alpha_d$}
\label{sec:alphad}
\begin{figure}
\includegraphics*[width=0.47\textwidth]{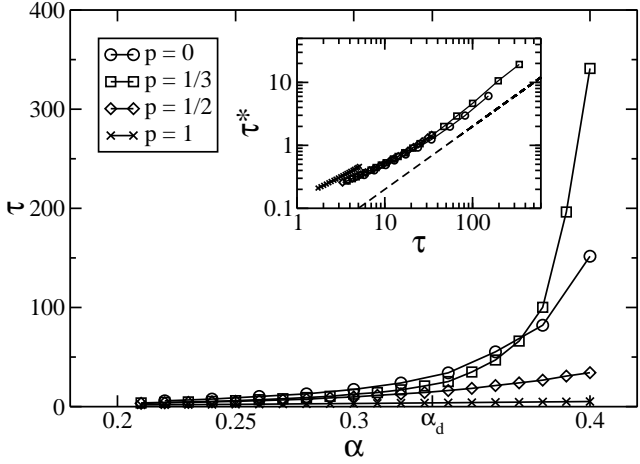}
\caption{Time $\tau$ for reaching a solution for a system of
$L=1000$ variables as a function of the ratio $\alpha$ and for
different values of the dynamical parameter $p$ [ $p=0$ (circles),
 $p=1/3$ (squares), $p=1/2$ (diamonds) and $p=1$ (crosses) ].
The $p$RWS performed for $p=1/3$ shows a critical behavior around
$\alpha_s=1/3$: for values of $\alpha \leq \alpha_s$, $\tau$ grows
almost linearly with $\alpha$, while it jumps to an exponential
growth with $\alpha$ for $\alpha> \alpha_s$. The same is
qualitatively true for $p=0$, but the time $\tau$ needed for
reaching a solution increases more slowly with respect to
the case $p=1/3$ for $\alpha > \alpha_s$. For $p=1/2$ and $p=1$
there seems to be no drastic increment of $\tau$ for $\alpha >
\alpha_s$. Moreover the inset shows the dependence of $\tau^*$,
the freezing time as calculated in the standard RWS \cite{weigt,semerjian},
on the freezing time $\tau$ calculated according to
Eq.(\ref{eq:time_scaling}). The almost linear dependence of
$\tau^*$ on $\tau$ (the dashed line has slope one) explains that
there is no qualitative change if we describe the dynamical
features of the system in terms of $\tau$ or $\tau^*$ as time used
by the simulations.} \label{fig:time_diluted}
\end{figure}

The freezing time $\tau$, that is the time
$\tau$ needed for finding a solution of the problem, abruptly
changes its behavior at the dynamical critical point $\alpha_d =
1/k$.
\\
Figure  \ref{fig:time_diluted} reports the numerical estimate of
the freezing time $\tau$ as a function of the dilution parameter
$\alpha$ and for different values of the dynamical parameter $p$ [
$p=0$ (circles) , $p=1/3$ (squares) , $p=1/2$ (diamonds) and $p=1$
(crosses) ]. As one can easily see, for $p=1/3$ and $p=0$, $\tau$
drastically changes around $\alpha_d$, increasing abruptly for
values of $\alpha> \alpha_d$. For $p=1/2$ and for $p=1$ this
drastic change is not observed. This is understandable from
the fact that both values of $p$ provide a bias towards paradise,
while $p=1/3$ corresponds to a random selection of one of the
three links of a triad as in the original RWS and $p=0$ would
favor the approach to the hell if it were a balanced state. The
simulations are performed over a system with $L=10^3$ variables.
Moreover each point stands for the average over $10^2$ different
networks and $10^2$ different realizations of the dynamics on such
topologies. At the beginning of each simulation the variables take
the value $1$ or $0$ with the same probability. The inset shows
the relation between the time $\tau^*$ calculated using the
standard RWS and the time $\tau$ calculated according to
Eq.(\ref{eq:time_scaling}). The almost linear relation (the dashed
line has a slope equal to one) between $\tau^*$ and $\tau$ means
that there is no qualitative change between the two different ways
of counting the time.

\begin{figure}
\includegraphics*[width=0.47\textwidth]{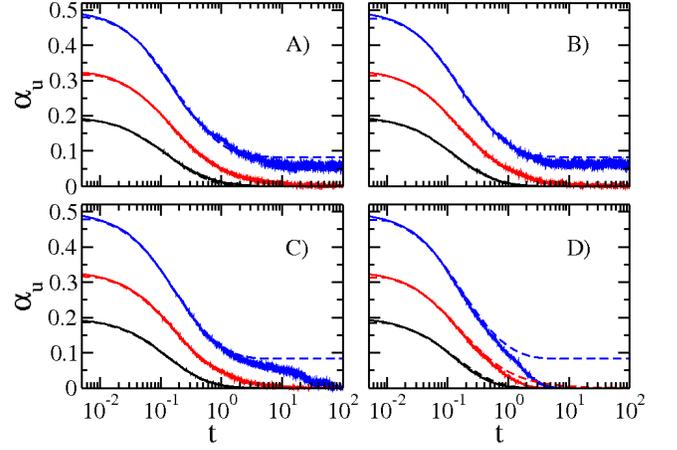}
\caption{(Color online) Time behavior of the ratio $\alpha_u$ of
unsatisfied clauses for different values of $p$ [ A) $p=0$ , B)
$p=1/3$ , C) $p=1/2$ , D) $p=1$ ] and for different values of the
dilution parameter $\alpha$ [ $\alpha=0.3$ (black, bottom) , $\alpha=0.5$
(red, middle) , $\alpha=0.85$ (blue, top) ]. Numerical results of simulations
[full lines] are compared with the numerical integration of
Eq.(\ref{eq:differential}) [dashed lines] leading to a very good
fit in all cases, except for $\alpha=0.85$ and for $p=1/2$ and
$p=1$. The initial configuration in all the cases is the one of an
antagonistic society ($x_i=0 \;\;\; , \; \forall \; i=1,\ldots ,L$), while
the number of variables is $L=10^4$.} \label{fig:timebehav}
\end{figure}

Following the same argument as in \cite{weigt}, we
can specify for the update event at time $t$ the variation of the
number of unsatisfied clauses $M_t^{(u)}$ as
\[
\Delta M_t^{(u)}=-\left(k \alpha_u(t) +1\right)+k \alpha_s(t) = k \alpha - 2 k \alpha_u(t)-1\;\;\;,
\]
because, by flipping one variable of an unsatisfied clause, all
the other unsatisfied clauses which share the same variable become
satisfied, while all the satisfied clauses containing that variable
become unsatisfied. In the thermodynamic limit $L\to \infty$, one
can impose $M_t^{(u)}=L\alpha_u(t)$. Moreover, the amount of time
of one update event is given by Eq.(\ref{eq:time_scaling}) so that
we can write
\begin{equation}
\dot{\alpha}_u(t)=\frac{\alpha_u(t)}{\alpha}\left(k\alpha-2k\alpha_u(t)-1\right)\;\;\;.
\label{eq:differential}
\end{equation}
Eq.(\ref{eq:differential}) has as stationary state (or a plateau) at
\begin{equation}
\alpha_u=\frac{k\alpha-1}{2k}\;\;\;. \label{eq:plateu}
\end{equation}
Therefore, when the ratio $\alpha$ (that is the ratio of the
number of clauses over the number of variables) exceeds the
critical ``dynamical'' value
\begin{equation}
\alpha_d = \frac{1}{k}\;\;\;,
\label{eq:weigt}
\end{equation}
the possibility of finding a solution for the problem drastically
changes. This result was already found by \cite{weigt,semerjian}. While for
values of $\alpha \leq \alpha_d$ we can always find a solution
because the plateau of Eq.(\ref{eq:plateu}) is always smaller or
equal to zero, for $\alpha>\alpha_d$ the solution is reachable
only if the system performs a fluctuation large enough to reach
zero from the non-zero plateau of Eq.(\ref{eq:plateu}). In Figure 
\ref{fig:timebehav} we report some numerical simulations for
$\alpha_u$ as a function of the time for different values of $p$ [
A) $p=0$ , B) $p=1/3$ , C) $p=1/2$ , D) $p=1$ ] and for different
values of the dilution parameter $\alpha$  [ $\alpha=0.3$ (black, bottom) , $\alpha=0.5$ (red, middle) , $\alpha=0.85$ (blue, top) ]. The numerical values
[full lines]  are compared with the numerical integration of
Eq.(\ref{eq:differential}) [dashed lines]. They fit very well
apart from large values of $t$, for $\alpha=0.85$ and for $p=1/2$
or $p=1$. The initial configuration in all cases is that of an
antagonistic society ($x_i=0 \;\;\; , \; \forall \; i=1,\ldots ,L$), while
the number of variables is $L=10^4$.

\subsubsection{Clustering of solutions at $\alpha_s$}
\label{sec:alphas}
\begin{figure}
\includegraphics*[width=0.47\textwidth]{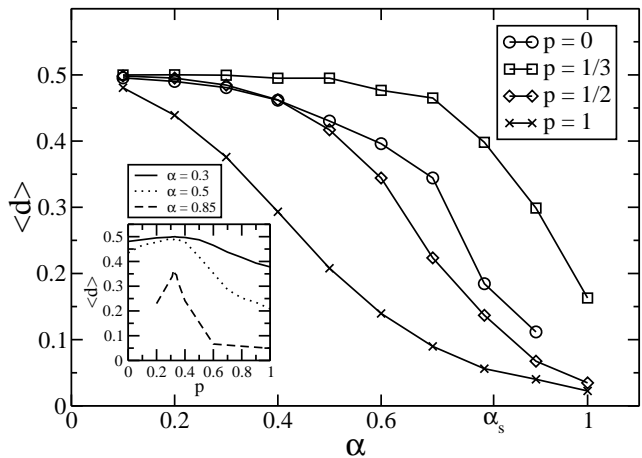}
\caption{Normalized Hamming distance $\langle d \rangle$ [
Eq.(\ref{eq:Hamming}) ] between solutions as a function of the
ratio $\alpha$ and different values of the dynamical parameter $p$
[ $p=0$ (circles) , $p=1/3$ (squares) , $p=1/2$ (diamonds) and
$p=1$ (crosses)]. For the standard RWS ($p=1/3$) the distance
drops down around the critical point $\alpha_s$. Different values
of $p$ perform not-really random walks and lead to effective
values of $\alpha_s$ smaller than the former one. The inset shows
the dependence of $\langle d \rangle$ on the dynamical parameter
$p$. As it is shown for different values of $\alpha$ [$\alpha=0.3$
(full line) , $\alpha=0.5$ (dotted line) and $\alpha=0.85$ (dashed
line)] the peak of the distance between solutions is for a $p$RWS
which is really random, that is for $p=1/3$. All the points here ,
in the main plot as well as in the inset, are obtained for a
system of $L=20$ variables. Each point is obtained averaging over
$10^2$ different networks and on each of these networks the
average distance is calculated over $10^2$ solutions. At the
beginning of each simulation the value of one variable is chosen to
be $1$ or $0$ with equal probability.} \label{fig:dist}
\end{figure}

In order to study the transition in the
clustering structure of solutions at $\alpha_s$, we numerically
determine the Hamming distance between different solutions of the
same problem. More precisely, given a problem of $L$ variables and
$M$ clauses, we find $T$ solutions $\left\{x_i^r \right\}_{i=1,\ldots ,L}^{r=1,\ldots , T}$ of the given problem. This
means that we start $T$ times from a random initial configuration
and at each time we perform a $p$RWS until we end up with a solution.
We then compute the distance between these $T$ solutions as
normalized Hamming distance
\begin{equation}
\langle d \rangle = \frac{1}{L\cdot T(T-1)}\sum_{r,s=1}^T
\sum_{i=1}^L \left| x_i^r - x_i^s \right| \;\;\;.
\label{eq:Hamming}
\end{equation}
The numerical results for $L=20$ are reported in Figure 
\ref{fig:dist}. We average the distance over $T=10^2$ trials and
over $10^2$ different problems for each value of $\alpha$. As
expected for $p=1/3$ [squares] the distance between solutions
drops down around $\alpha_s$ (actually it drops down before
$\alpha_s$ because of the small number of variables). For
different values of $p$ [ $p=0$ (circles) , $p=1/2$ (diamonds) and
$p=1$ (crosses) ], the $p$RWS is less random and $\langle d \rangle$
drops down before $\alpha_s$ (or at least before the point at
which the case $p=1/3$ drops down). In particular, if we plot (as
in the inset) the distance $\langle d \rangle$ as a function of
$p$ and for different values of $\alpha$ [$\alpha=0.3$ (full line)
, $\alpha=0.5$ (dotted line) and $\alpha=0.85$ (dashed line)] we
see a clear peak of the distance $\langle d \rangle$ around
$p=1/3$. This suggests that a completely random, unbiased RWS
always explores a large region in phase space, it leads to a
larger variety of solutions.

\subsubsection{SAT/UNSAT transition at $\alpha_c$}
\label{sec:pc}
\begin{figure}
\includegraphics*[width=0.47\textwidth]{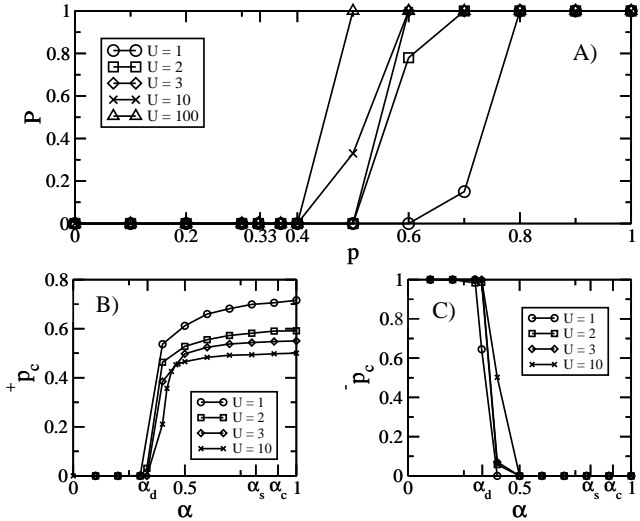}
\caption{Numerical estimate of the upper  $^+p_c$ (B) and lower
$^-p_c$ (C) critical values of $p$ [see the text for their
definition] as a function of the dilution parameter $\alpha$. Here
$L=10^3$ and the different symbols corresponds to different
maximum waiting times $T=U\cdot L$ [ $U=1$ (circles) ,  $U=2$
(squares) ,   $U=3$ (diamonds) ,  $U=10$ (diamonds) ]. Each point
is given by the average over  $10$ different problems for each
value of $\alpha$ and $10$ different $p$RWS for each problem
(with random initial condition). Moreover in (A) we show the
probability $P$ that the $p$RWS finds a solution at $\alpha
=0.918 \simeq \alpha_c$ as a function of $p$. We used different
waiting times [ $U=1$ (circles) ,  $U=2$ (squares) , $U=3$
(diamonds) , $U=10$ (crosses) , $U=100$ (triangles) ]. See the
text for further comments.} \label{fig:pcritic}
\end{figure}

Differently from the general $k$S problem, the
$k$XS problem is known to be always solvable
\cite{semerjian} and the solution corresponds to one of the
balanced configurations as described in section \ref{frozen} for
the all-to-all topology. Nevertheless the challenge is whether
the solutions can be found by a local random algorithm like RWS.
In the application of the RWS it can happen that the algorithm is
not able to find one of these solutions in a ``finite'' time, so
that the problem is called ``unsatisfied''. The notion is made
more precise in \cite{cocco}. For practical reasons the way of
estimating the critical point $\alpha_c$ that separates the SAT
from the UNSAT region is related to the so-called algorithm complexity of
the RWS. Here we follow the prescription of
\cite{weigt,semerjian,schoning}. Fixed $k=3$ and calling a RWS with initial random assignment of
the variables followed by $3L$ update events one trial, one needs
a total number of trials $T \gg \left(4/3\right)^L$ for being
``numerically'' sure to be in the UNSAT region. In fact if after
$T$ trials  no solution is found, the problem is considered as
``unsatisfied'' .
\\
The introduction of the dynamical parameter $p$ can strongly
``improve'' the performance of RWS. For $p \neq 1/3$ the $p$RWS
updates the variables following a well prescribed direction: the
tendency is to increases the number of negative variables for
$p<1/3$ and to decrease their number for $p > 1/3$. In particular,
as we have seen in the former sections, for $p \geq 1/2$ the $p$RWS
approaches the configuration of the paradise
for the largest value of $\alpha={L \choose k}/L \gg \alpha_c$ and
in a time that goes as $\tau \sim L^\beta$,  so that there is no
UNSAT region at all if we apply the former criterion for the
numerical estimate of the UNSAT region. Clearly, if the bias goes
in the wrong direction, the performance gets worse.
\\
In this section we briefly give a qualitative description about
the SAT/UNSAT region for the $p$RWS due to the dynamical
parameter $p$. Let us define as $^+p_c$ [$^-p_c$] the minimum
[maximum] value of $p$ for which the system can be satisfied.
Given a problem with $\alpha L$ clauses we follow the algorithm:
1)  Set $p=1$ [$p=0$] ; 2) set an initial random configuration and
apply the $p$RWS ; 3) if the $p$RWS finds the solution in a number of
updates less than $U\cdot L$ , decrease [increase] $p$ and go to
point 2) ; 4) if not $^+p_c=p$ [$^-p_c=p$]. This procedure can be
performed up to the desired sensitivity for the numerical estimate
of $^+p_c$ [$^-p_c$]. The idea of defining an upper $^+p_c$ and
lower critical value $^-p_c$ for the dynamical parameter $p$ is
related to the fact that for $p=1/3$ the $p$RWS has most trouble to
find the solution. Figure  \ref{fig:pcritic}B and Figure 
\ref{fig:pcritic}C show the numerical results for $^+p_c$ and
$^-p_c$ as a function of the dilution parameter $\alpha$. The
number of variables is $L=10^3$. We report the results for
different values of the waiting time $T=L \cdot U$ [ $U=1$
(circles) ,  $U=2$ (squares) ,   $U=3$ (crosses) ,  $U=10$
(crosses) ]. Each point is averaged over $10$ different problems
and $10$ different $p$RWS applied to each problem. Qualitatively it
is seen that for $\alpha \leq \alpha_d$ the problem is always
solvable ( $^+p_c=0$ and $^-p_c=1$ ) , while for $\alpha >
\alpha_d$ one needs $p \neq 1/3$ for solving the problem. Of
course the numerical values for $^+p_c$ and $^-p_c$ depend on the
waiting time until the $p$RWS reaches a solution. Here, for
simplicity we do not wait long enough for seeing a similar
behavior around $\alpha_c$ instead of $\alpha_d$. Furthermore, in
Figure  \ref{fig:pcritic}A we report the probability $P$, that is
the ratio of success over the number of trials, for solving the
problem as a function of $p$ for $\alpha=\alpha_c$. The waiting
time is  $U=1$ (circles) ,  $U=2$ (squares) ,   $U=3$ (diamonds) ,
$U=10$ (crosses) and $U=100$ (triangles), respectively. The
probabilities are calculated over $10^2$ trials for each point
($10$ different problems times $10$ $p$RWS for each problem). As the
waiting time increases the upper critical value $^+p_c$ for
finding for sure the solution decreases ( $^+p_c \simeq 0.8$ for
$U=1$ , $^+p_c \simeq 0.7$ for $U=2$ , $^+p_c \simeq 0.6$ for
$U=3$ and for $U=10$ , $^+p_c \simeq 0.5$ for $U=100$ ). This
means that even for less biased search, solutions can be found,
while $^-p_c$ is zero for the waiting time reported here, no value
of $p<p_c$ leads to a solution. This is as expected. If the
variables are almost all negative it is harder to find a solution
of the problem (the paradise is a solution while the hell for $k$
odd is not).

\subsubsection{Mean-field approximation down to $\alpha_m$}
\label{sec:alpham}

\begin{figure}
\includegraphics*[width=0.47\textwidth]{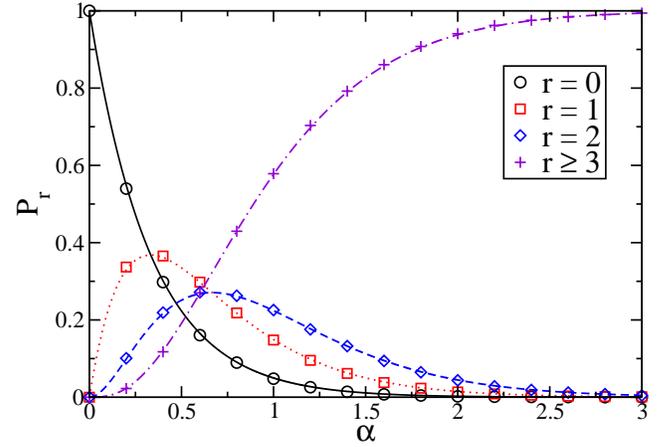}
\caption{Probability $p_r$ that one variable belongs to $r$
clauses as function of the dilution parameter $\alpha$. The
symbols stand for numerical results obtained over $10^3$ different
realizations for $L=128$ variables [ $r=0$ (black circles) , $r=1$
(red squares) , $r=2$ (blue diamonds) and $r \geq 3$ (violet
crosses) ]. The lines stand for analytical predictions of
Eq.(\ref{eq:probdil}) [ $r=0$ (black full line) , $r=1$ (red
dotted line) , $r=2$ (blue dashed line) and $r \geq 3$ (violet
dotted-dashed line)  ].} \label{fig:probdil}
\end{figure}

By construction the ``topology'' of a $k$S problem  is completely random (for this reason is sometimes called explicitly as Random $k$-SAT problem). Each of the $L$ variables can appear in one
of the $\alpha L$ clauses with probability
$v=\frac{1}{L}+\frac{1}{L-1}+\ldots + \frac{1}{L-k}$. In
particular for $L \gg k$ one can simply write $v\simeq
\frac{k}{L}$. Then the probability $P_r$ that one variable belongs
to $r$ clauses can be described by the Poisson distribution
\begin{equation}
P_r = \frac{\left(\alpha k \right)^r}{r!} e^{-\alpha k} \;\;\; ,
\label{eq:probdil}
\end{equation}
with mean value $\langle r \rangle = \alpha k$ and variance
$\sigma_r = \sqrt{\alpha k}$. $P_r$ is plotted in Figure 
\ref{fig:probdil}, where the numerical results [ symbols ,  $r=0$
(black circles) , $r=1$ (red squares) , $r=2$ (blue diamonds) and
$r \geq 3$ (violet crosses) ] are compared to the analytical
expectation [ lines , $r=0$ (black full line) , $r=1$ (red dotted
line) , $r=2$ (blue dashed line) and $r \geq 3$ (violet
dotted-dashed line)  ].

\begin{figure}
\includegraphics*[width=0.47\textwidth]{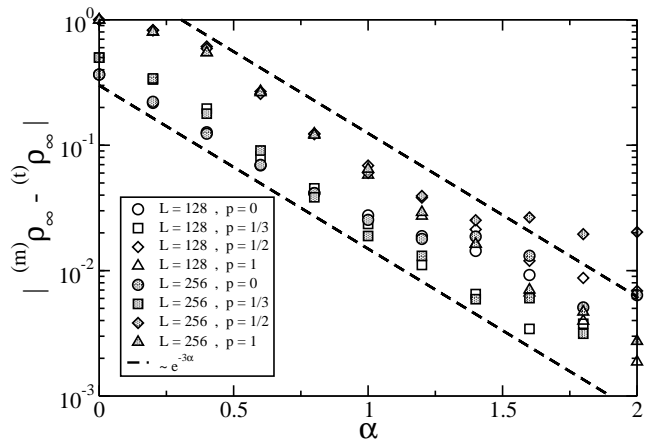}
\caption{Difference $\left| \; ^{(m)}\rho_\infty - \;
^{(t)}\rho_\infty \; \right| $  between $^{(t)}\rho_\infty$ the
theoretical prediction for the stationary density of friendly
variables [ Eq.(\ref{eq:antal2}) ] and the numerically measured
value  $^{(m)}\rho_\infty$, as a function of the dilution
parameter $\alpha$. $^{(m)}\rho_\infty$ is obtained as the average
of the density of friendly links (registered after a waiting time
$T=200.0$ , so that is effectively stationary ) over $50$
different problems and $50$ different $p$RWS for each problem.
The results displayed here are obtained for $L=128$ (open symbols)
and $L=256$ (gray filled symbols) and for different values of $p$
[ $p=0$ (circles) , $p=1/3$ (squares) , $p=1/2$ (diamonds) , $p=1$
(triangles) ]. The initial conditions are those of an antagonistic
society. The dashed lines are proportional to $e^{-3 \alpha}$.}
\label{fig:dendil}
\end{figure}

If we start from an antagonistic society (all variables false)
the minimum value of the dilution  $\alpha_m$ needed to reach  the
paradise (if $p \geq 1/2$) is that all variables belong to at
least one clause. This means that $P_0 < 1/L$, from which
\begin{equation}
\alpha_m = \frac{\ln{L}}{k} \;\;\; .
\label{eq:meanfield}
\end{equation}
 It is interesting to note that the same criterion applies for any $p$.
 In Figure  \ref{fig:dendil} we plot the absolute value of the
 difference $\left| \; ^{(m)}\rho_\infty - \; ^{(t)}\rho_\infty \; \right| $,
 between $^{(t)}\rho_\infty$, the theoretical prediction for the
 stationary density of true variables, [ Eq.(\ref{eq:antal2}) ]
 and the numerically measured value  $^{(m)}\rho_\infty$, as a
 function of the dilution parameter $\alpha$. $^{(m)}\rho_\infty$ is
 obtained as the average of the density of friendly links
 (registered after a waiting time $T=200.0$ , so that is effectively the stationary density)
 over $50$ different problems and $50$ different $p$RWS for each problem.
 The results reported here are for $L=128$ (open symbols) and $L=256$
 (gray filled symbols) and for different values of $p$ [ $p=0$ (circles) ,
 $p=1/3$ (squares) , $p=1/2$ (diamonds) , $p=1$ (triangles) ].
 The initial conditions are those of an antagonistic society.
 The dashed lines are proportional to $e^{-3 \alpha}$. Figure  \ref{fig:dendil}
 shows that the mean-field approximation of Eq.(\ref{eq:antal2})
 becomes exponentially fast true as the system dilution decreases.
 Moreover, as for the cases $p=0$ and $p=1/3$, we can observe that
 the difference $\left| \; ^{(m)}\rho_\infty - \; ^{(t)}\rho_\infty \; \right| $
 is always smaller than for $p=1/2$ and $p=1$.
 Qualitatively this means that the dilution $\alpha$ of
 the system needed to reach the theoretical expectation of
 Eq.(\ref{eq:antal2}) is smaller than $\alpha_m$ for $p<1/2$.
 In general we can say that $\alpha_m$ is a function of $p$:
 $\alpha_m=\alpha_m(p)$, and $\alpha_m$ is the minimum value
 of the dilution of the system for which we can effectively describe
 the diluted system as an all-to-all system for all the values of
 $p$.
 Moreover, it should be noticed that for $\alpha > \alpha_m$ almost
 all variables belong to at least three clauses [ see Figure  \ref{fig:probdil}
 ]. This fact allows the $p$RWS to explore a larger part of configuration
 space.
 Let us assume that one variable belongs to less than three clauses: an eventual update event
 that flips this variable so that the one triad becomes balanced, can never increase
 the number of unsatisfied clauses by frustrating
 other clauses it belongs to. This reminds us to the
 situation in an energy landscape in which an algorithm gets stuck
 in a local minimum when it never accepts a change in the ``wrong''
 direction, i.e. towards larger energy.

\section{Summary and conclusions}\label{summary}
In the first part of this paper we generalized the triad dynamics
of Antal \emph{et al.} to a $k$-cycle dynamics \cite{antal}. Here we had to
distinguish the cases of even values of $k$ and odd values of $k$. For all values of
integer $k$ there is again a critical threshold at $p_c=1/2$ in
the propensity parameter. For odd $k$ and $p<p_c$ the paradise can
never be reached in the thermodynamic limit of infinite
system size (as predicted by the mean field equations which we
solved exactly for $k=5$ and approximately for $k>5$). In the finite volume, in principle one could reach
a balanced state made out of two cliques (a special case of this configuration is the ``paradise'' when one clique is empty). However, the probability
for reaching such type of frozen state decreases exponentially
with the system size so that in practice the fluctuations never
die out in the numerical simulations. For $p>1/2$ the convergence
time to reach the paradise grows logarithmically with the system
size. At $p=1/2$ paradise is reached within a time that follows a
power law in the size $N$, where we determined the $k$-dependence
of the exponent. In particular, the densities of $k$-cycles with $j$ negative links,
here evolved according to the rules of the $k$-cycle dynamics,
could be equally well obtained from a random dynamics in which
each link is set equal to $1$ with probability $\rho_\infty$ or
equal to $-1$ with probability $1-\rho_\infty$. This feature was
already observed by Antal \emph{et al.} for $k=3$ \cite{antal}. It means that the
individual updating rules which seem to be ``socially'' motivated in
locally reducing the social tensions by changing links to friendly
ones, end up with random distributions of friendly links. The
reason is a missing constraint of the type that the overall number
of frustrated $k$-cycles should not increase in an update event.
Such a constrained dynamics was studied by Antal \emph{et al.} in
\cite{antal}, but not in this paper.
\\
For even values of $k$, the only stable solutions are
``heaven'' (i.e. paradise) and ``hell'' for $p>1/2$ and $p<1/2$,
respectively, and the time to reach these frozen configurations
grows logarithmically with $N$. At $p_c=1/2$ other realizations of
the frozen configurations are possible, in principle. However, they have negligible
probability as compared to heaven and hell. Here the time to reach
these configurations increases quadratically in $N$, independently
of $k$. This result was obtained in two ways. Either from the
criterion to reach the stable state when a large enough
fluctuation drops the system into this state (so we had to
calculate how long one has to wait for such a big fluctuation).
Alternatively, the result could be read off from a mapping to a
Markov process for diploid organisms ending up in a genetic pool
of either all ``$+$''-genes or all ``$-$''-genes. The difference in the
possible stable states of diploid organisms and ours consists in
two-clique stable solutions that are admissible for the even
$k$-cycle dynamics, in principle, however such clique states have
such a low probability of being realized that the difference is
irrelevant.
\\
The difference in the exponent at $p_c$ and the stable
configurations above and below $p_c$ between the even and odd
$k$-cycle dynamics was due to the fact that "hell", a state with
all links negative as in an antagonistic society, is a balanced
state for even $k$, not only by the frustration criterion of
physicists, but also according to the criterion of social
scientists \cite{cartwright}.
\\
\vspace{5pt}
\\
As a second natural generalization of the social balance
dynamics of Antal \emph{et al.} we considered a diluted network. One way of implementing the
dilution is via a random Erd\"os-R\'enyi network, characterized by
the probability $w$ for connecting a randomly chosen pair of
nodes. Here we focused our studies to the case $k=3$.
The mean-field description and the results about the phase
structure remain valid down to a certain degree of dilution,
characterized by $w_m$. This threshold for the validity of the
mean-field description practically coincides with the criterion
whether a single link belongs to at least three triads (for
$w>w_m$) or not ($w<w_m$). If it does so, an update event can
increase the number of frustrated triads. For $w<w_m$, or more
precisely $w<w_d<w_m$ it becomes easier to realize frozen
configurations different from the paradise. Isolated links do not
get updated at all and isolated triads can freeze to a
``$+$''-``$-$''-configuration. The time to reach such a frozen configuration
(in general different from the paradise) grows then only linearly
in the system size. Also the solution space, characterized by the
average Hamming distance between solutions, has different features
below and above another threshold, called $w_s$ with
$w_d<w_s<w_m$.
Therefore one of the main differences between the all-to-all and
the sufficiently diluted topology are the frozen configurations.
For the all-to-all case we observed the paradise above $p_c$ for
odd values of $k$ and even values of $k$ and the hell for even values of $k$ below $p_c$, in the numerical simulations, because the
probability to find a two-clique-frozen configuration was
calculated to be negligibly small. For larger dilution, also other
balanced configurations were numerically found, as mentioned
above, and the time passed in the numerical simulations for
finding these solutions followed the theoretical predictions.
\\
In section \ref{sec:diluted} we used, however, another parameterization in
terms of the dilution parameter $\alpha$, that was the ratio of
triads (clauses) over the number of links [we gave anyway an approximated relation between $\alpha$ and $w$ in Eq.(\ref{eq:average_ratio})]. The reason for using
this parameterization was a mapping of the $k$-cycle social
balance of networks to a $k$-XOR-SAT ($k$XS) problem, that is a typical
satisfiability problem in optimization tasks. We also traced a mapping between the
``social'' dynamical rules and the  Random-Walk SAT (RWS) algorithm, that is one
approach for solving this problem in a random local way. As we
have shown, the diluted version of the $3$-cycle social dynamics
with propensity parameter $p=1/3$ corresponds to a
$3$XS problem solved by the RWS algorithm in its
standard form (as used in \cite{weigt,semerjian}).
\\
The $k$XS problem is always solvable like the $k$-cycle
social balance, for which a two cliques solution always exists due
to the structure theorem of \cite{cartwright}, containing as a
special solution the so-called paradise. The common challenge,
however, is to find this solution by a local stochastic algorithm. The driving force, shared by both sets of
problems, is the reduction of frustration. The meaning of
frustration depends on the context: for the $k$-cycle dynamics it
is meant in a social sense as a reduction of social tension, for
the $k$XS problem it corresponds to violated clauses.
The mathematical criterion is the same. The local stochastic
algorithm works in a certain parameter range, but outside this
range it fails. The paradise is never reached for a propensity
parameter $p<1/2$, independently of $k$. Similarly, the solution
of the $k$XS problem is never found if the dilution
parameter is larger than $\alpha_c$, and the RWS algorithm
needs an exponentially long time already for $\alpha>\alpha_d$,
with $\alpha_d<\alpha_c$.
\\
We generalized the RWS algorithm, usually chosen for
solving the $k$-SAT ($k$S) problem as well as the
$k$XS problem, to include a parameter $p$ that played
formerly the role of the propensity parameter in the social
dynamics ($p$RWS). The effect of this parameter is a bias towards the
solution so that $\alpha_d$, the threshold between a linear and an
exponential time for solving the problem, becomes a function of
$p$. Problems for which the $p$RWS algorithm needed
exponentially long for $p=1/3$, now become solvable within a time that grows less than
logarithmically in the system size for $p>1/2$ and less than power-like  in the system size for $p=1/2$.  Along with the
bias goes an exploration of solution space that has on average a
smaller Hamming distance between different solutions than in the
case of the $\frac{1}{3}$RWS algorithm that was formerly
considered \cite{weigt,semerjian}.
\\
\vspace{5pt}
\\
Our paper has illustrated that the reduction of
frustration may be the driving force in common to a number of
dynamical systems. So far we were concerned about ``artificial''
systems like social systems and satisfiability problems. It would
be interesting to search for natural networks whose evolution was
determined by the goal of reducing the frustration, not
necessarily to zero degree, but to a low degree at least.

\begin{acknowledgments}
It is a pleasure to thank Martin Weigt for drawing our attention
to Random $k$-SAT problems in computer science and for having  useful discussions with us while he was visiting the International University Bremen as an
ICTS-fellow.
\end{acknowledgments}

\end{document}